\documentclass[12pt]{article}

\usepackage[T1]{fontenc}
\usepackage[utf8x]{inputenc}

\usepackage{authblk}
\usepackage{fullpage}
\usepackage{amssymb,amsmath}
\usepackage{siunitx}
\usepackage[version=3]{mhchem}
\usepackage{mathptmx}
\usepackage{natbib}
\usepackage{aligned-overset}
\usepackage[dvipsnames]{xcolor} 
\usepackage{bm}
\usepackage{multirow}
\usepackage{url}
\usepackage{ulem}
\usepackage{pdflscape}

\usepackage[left]{lineno}
\usepackage{setspace}
\usepackage{graphicx}
\usepackage{multirow}
\usepackage{colortbl}

\newcommand{\bzero}{\mbox{\boldmath $0$}}

\definecolor{Gray}{gray}{0.9}

\makeatletter
\def\ScaleWidthIfNeeded{%
 \ifdim\Gin@nat@width>\linewidth
    \linewidth
  \else
    \Gin@nat@width
  \fi
}
\def\ScaleHeightIfNeeded{%
  \ifdim\Gin@nat@height>0.9\textheight
    0.9\textheight
  \else
    \Gin@nat@width
  \fi
}
\makeatother
\setkeys{Gin}{width=\ScaleWidthIfNeeded,height=\ScaleHeightIfNeeded,keepaspectratio}%

\title{A General Statistical Framework for Hardy–Weinberg Equilibrium Inference on the X Chromosome}
\author[1]{Lin Zhang}
\author[2,3,4]{Andrew D. Paterson}
\author[5,4]{Lei Sun}
\affil[1]{Department of Statistics and Actuarial Science, Faculty of Science, Simon Fraser University, Burnaby, British Columbia, Canada\\ lin\_zhang\_7@sfu.ca}
\affil[2]{Genetics and Genome Biology, The Hospital for Sick Children, Toronto, Ontario, Canada\\ andrew.paterson@sickkids.ca}
\affil[3]{Epidemiology Division, Dalla Lana School of Public Health, University of Toronto, Toronto, Ontario, Canada}
\affil[4]{Biostatistics Division, Dalla Lana School of Public Health, University of Toronto, Toronto, Ontario, Canada}
\affil[5]{Department of Statistical Sciences, Faculty of Arts and Science, University of Toronto, Toronto, Ontario, Canada\\ lei.sun@utoronto.ca}

\date{\today}

\begin{document}
\maketitle
\newpage

%%%%%%%%%%%%%%%%%%%%%%%%%%%%%%%%%%%%%%%%%%%%%%%%%%%%%%%%%%%%%%%%%%%%%%%%%%%%%%
%%%%%%%%%%%%%%%%%%%%%%%%%%%%%%%%%%%%%%%%%%%%%%%%%%%%%%%%%%%%%%%%%%%%%%%%%%%%%%
%%%%%%%%%%%%%%%%%%%%%%%%%%%%%%%%%%%%%%%%%%%%%%%%%%%%%%%%%%%%%%%%%%%%%%%%%%%%%%
\section*{Abstract}

Testing for Hardy--Weinberg equilibrium (HWE) is a fundamental component of genetic data analysis, widely used for quality control and model validation. Although HWE testing is well established for autosomal loci, inference on the X chromosome is more complex due to sex-specific genotype structures and potential sex differences in minor allele frequency (sdMAF). Existing tests differ in their assumptions about sdMAF and male sample inclusion, often leading to distinct but poorly characterized null hypotheses.

We develop a general statistical framework for HWE inference using the robust allele-based regression model. By formulating HWE testing as an assessment of allele-level dependence, the framework directly parameterizes Hardy--Weinberg disequilibrium, unifies existing Pearson $\chi^2$-based tests under explicit modeling assumptions, and clarifies their null hypotheses, degrees of freedom, and sensitivity to sdMAF. The framework also accommodates covariate and population-structure adjustment within a unified regression-based formulation.

The proposed framework provides robust, interpretable, and flexible inference, establishing a unified statistical foundation for HWE testing across autosomal and X-chromosomal regions. Simulation studies and analysis of high-coverage 1000 Genomes Project data demonstrate that commonly used X-chromosome tests can exhibit inflated type I error or misleading inference when sdMAF is present.

%%%%%%%%%%%%%%%%%%%%%%%%%%%%%%%%%%%%%%%%%%%%%%%%%%%%%%%%%%%%%%%%%%%%%%%%%%%%%%
%%%%%%%%%%%%%%%%%%%%%%%%%%%%%%%%%%%%%%%%%%%%%%%%%%%%%%%%%%%%%%%%%%%%%%%%%%%%%%
%%%%%%%%%%%%%%%%%%%%%%%%%%%%%%%%%%%%%%%%%%%%%%%%%%%%%%%%%%%%%%%%%%%%%%%%%%%%%%

\bigskip
\bigskip
\noindent
{\it Keywords}: Hardy–Weinberg equilibrium; X chromosome; sex difference in minor allele frequency; robust allele-based regression; score test; Pearson chi-square test
\newpage

%%%%%%%%%%%%%%%%%%%%%%%%%%%%%%%%%%%%%%%%%%%%%%%%%%%%%%%%%%%%%%%%%%%%%%%%%%%%%%
%%%%%%%%%%%%%%%%%%%%%%%%%%%%%%%%%%%%%%%%%%%%%%%%%%%%%%%%%%%%%%%%%%%%%%%%%%%%%%
%%%%%%%%%%%%%%%%%%%%%%%%%%%%%%%%%%%%%%%%%%%%%%%%%%%%%%%%%%%%%%%%%%%%%%%%%%%%%%
\section{Introduction} \label{s:intro}

The concept of Hardy--Weinberg equilibrium (HWE) was originally introduced for an autosomal single-nucleotide polymorphism (SNP), where HWE implies independent coupling of the two alleles in a genotype \citep{hardy1908mendelian, weinberg1908demonstration, crow1970introduction, olson1996testing}. Testing for deviation from HWE, or Hardy–Weinberg disequilibrium (HWD), has become a fundamental component of genetic data analysis, widely used for quality control, model validation, and detection of departures from core population-genetic assumptions.

For an autosomal SNP, HWE testing compares the observed genotype frequencies with those expected under the null of HWE, calculated using a single allele frequency assumed for all individuals. In this setting, the null hypothesis is well defined and widely understood. In contrast, the HWE testing on the X chromosome (Xchr) is substantially more complex because females are diploid and males are hemizygous in the non-pseudoautosomal region (NPR), genotype distributions differ by sex, and the relationship between allele frequencies and genotype frequencies is no longer symmetric. In the pseudoautosomal regions (PAR), both sexes are effectively diploid, creating region-specific genotype structures that exclude a single classical formulation.

An additional complication arises from the possibility that allele frequencies differ between sexes \citep{wang2024population}. Such sex differences in minor allele frequency (sdMAF) can result from biological processes, including sex-biased admixture or selection, or from technical artifacts. When sdMAF is present, pooling data between sexes can induce apparent deviations from HWE even in the absence of a true disequilibrium or genotyping errors. Consequently, observed HWD may reflect sdMAF rather than true equilibrium violations, implying that the null hypothesis for Xchr HWE is not uniquely defined and depends critically on assumptions regarding allele frequencies and genotype structures.

Existing approaches either restrict analysis to female samples, treating the problem analogously to autosomal loci, or combine male and female samples under the pooled allele-frequency assumption. For example, \citet{graffelman2016testing} proposed a $\chi^2_2$ test that combines data from both sexes and compares the observed genotype counts to the expectations calculated based on the pooled allele frequencies between the sexes. Although widely used in genome-wide association studies (GWAS) involving the X chromosome, the underlying assumptions and null hypotheses of these approaches remain poorly understood.

A central yet often under-recognized issue is that commonly used Xchr HWE tests do not represent alternative implementations of a single inferential procedure; rather, they target distinct null hypotheses by implicitly or explicitly imposing different assumptions regarding allele frequencies, genotype structure, and the presence or absence of sdMAF. Consequently, Xchr HWE testing is fundamentally a model-specification problem, where valid and interpretable inference depends critically on making these assumptions explicit. A unified statistical framework is therefore needed to clarify the relationships among existing tests, distinguish their inferential targets, and enable more flexible analysis in complex genetic settings.

In this paper, we develop a general statistical framework for Xchr HWE inference by extending the robust allele-based (RA) regression framework that we previously introduced for allelic association testing \citep{zhang2022generalized}, which provides a natural regression-based formulation for allele-level dependence. The RA framework explicitly models allele dependence within individuals, thereby enabling a direct parameterization of HWD. Within this RA framework, existing HWE tests can be expressed as special cases corresponding to specific constraints on model parameters, including assumptions about sdMAF.

The resulting framework provides several important inferential advantages. First, it yields a unified interpretation of commonly used Pearson $\chi^2$-based tests by linking them to explicit statistical models. Second, it allows flexible extensions, including the incorporation of covariates and adjustment for population structure. Third, it separates the contributions of HWD and sdMAF, allowing a clearer interpretation of the test results when allele frequencies differ between sexes.

We clarify the proposed framework through theoretical derivations, simulation studies that evaluate type I error and power, and applications to high-coverage whole-genome sequence data from the 1000 Genomes Project \citep{byrska2022high}. Our findings demonstrate that explicit modeling assumptions are essential for valid, interpretable, and flexible HWE inference on the X chromosome.

%%%%%%%%%%%%%%%%%%%%%%%%%%%%%%%%%%%%%%%%%%%%%%%%%%%%%%%%%%%%%%%%%%%%%%%%%%%%%%
%%%%%%%%%%%%%%%%%%%%%%%%%%%%%%%%%%%%%%%%%%%%%%%%%%%%%%%%%%%%%%%%%%%%%%%%%%%%%%
%%%%%%%%%%%%%%%%%%%%%%%%%%%%%%%%%%%%%%%%%%%%%%%%%%%%%%%%%%%%%%%%%%%%%%%%%%%%%%
\section{Pearson $\chi^2$-based HWE tests: existing and derived forms}
\label{s:overview}

In this section, we review and extend Pearson $\chi^2$-based formulations for HWE testing for autosomal and X-chromosomal SNPs under biologically relevant assumptions regarding genotype structures, sex-specific allele frequencies, and the potential presence of sdMAF. These formulations establish the foundation for the classical HWE testing, provide important baseline procedures, and motivate the broader regression-based framework developed in subsequent sections.

We begin with the classical autosomal setting, which serves as the foundation for later extensions, and then consider X-chromosomal NPR and PAR SNPs, where male hemizygosity, sex-specific genotype structure, and sdMAF introduce additional complexity. By organizing these Pearson-based formulations under explicit assumptions, we clarify their relationships while preserving the classical goodness-of-fit perspective that underlies many widely used procedures.

%%%%%%%%%%%%%%%%%%%%%%%%%%%%%%%%%%%%%%%%%%%%%%%%%%%%%%%%%%%%%%%%%%%%%%%%%%%%%%
%%%%%%%%%%%%%%%%%%%%%%%%%%%%%%%%%%%%%%%%%%%%%%%%%%%%%%%%%%%%%%%%%%%%%%%%%%%%%%
\subsection{Notation} 
\label{ss:notations}

Consider a bi-allelic SNP with alleles denoted $a$ and $A$, with $p$ denoting the population frequency of allele $A$. Table~\ref{tab:notations} summarizes genotype counts and sample allele-frequency estimators, stratified by sex and genomic region (autosomal/Xchr PAR versus Xchr NPR).

For Xchr NPR SNPs, male genotypes are hemizygous ($a$ or $A$). We retain conventional subscripts $0$ and $2$ for male counts for notational consistency, with the estimate of male allele-frequency $\hat p_m=m_2/m$, in contrast to the diploid estimator $\hat p_m=(m_1+2m_2)/(2m)$ used for autosomal and Xchr PAR SNPs. Throughout, the allele $A$ is defined relative to the female sample, allowing direct assessment of sdMAF between sexes.

\begin{table}[t!]
\begin{center}
\begin{tabular}{l|ccc|c|c}
\hline \hline
\multicolumn{6}{c}{(I) {\bf Autosomes} or {\bf Xchr PAR}} \\ \hline \hline
& \multicolumn{4}{c|}{Genotype Counts} & Allele Frequency Estimator for Allele $A$ \\ \hline
\multirow{2}{*}{Female} & $aa$ & $Aa$ & $AA$ & Total & \\ \cline{2-6} 
& $f_0$ & $f_1$ & $f_2$ & $f$ & $\hat p_f={(f_1+2f_2)}/{2f}$ \\ \hline
\multirow{2}{*}{Male} & $aa$ & $Aa$ & $AA$ & Total &  \\ \cline{2-6} 
& $m_0$ & $m_1$ & $m_2$ & $m$ & $\hat p_m={(m_1+2m_2)}/{2m}$ \\ \hline
Total & $n_0$ & $n_1$ & $n_2$ & $n$ & $\hat p={(f_1+2f_2+m_1 + 2m_2)}/{2n}={(n_1+2n_2)}/{2n}$  \\ \hline \hline
\multicolumn{6}{c}{(II) {\bf Xchr NPR}} \\ \hline \hline
& \multicolumn{4}{c|}{Genotype Counts} & Allele Frequency Estimator \\ \hline
\multirow{2}{*}{Female} & $aa$ & $Aa$ & $AA$ & Total &  \\ \cline{2-6}
 & $f_0$ & $f_1$ & $f_2$ & $f$ & $\hat p_f={(f_1+2f_2)}/{2f}$ \\ \hline
 \multirow{2}{*}{Male} & $a$ & $\ast$ & $A$ & Total &   \\ \cline{2-6}
 & $m_0$ & $\ast$ & $m_2$ & $m$ & $\hat p_m = {m_2}/{m}$\\  \hline
Total & $n_0$ & $n_1$ & $n_2$ & $n$ & $\hat p ={(f_1+2f_2+m_2)}/{(2f+m)}$ \\ \hline \hline 
\end{tabular}
\caption{{\bf Genotype counts and allele-frequency estimators for a bi-allelic SNP stratified by sex and genomic region.} Xchr PAR denotes the pseudoautosomal regions, and Xchr NPR denotes the non-pseudoautosomal regions. For Xchr NPR SNPs, male genotypes are hemizygous.}
\label{tab:notations}
\end{center}
\end{table}

%%%%%%%%%%%%%%%%%%%%%%%%%%%%%%%%%%%%%%%%%%%%%%%%%%%%%%%%%%%%%%%%%%%%%%%%%%%%%%
%%%%%%%%%%%%%%%%%%%%%%%%%%%%%%%%%%%%%%%%%%%%%%%%%%%%%%%%%%%%%%%%%%%%%%%%%%%%%%
\subsection{The classical Pearson $\chi^2_1$ and $\chi^2_2$ tests of HWE for autosomal or Xchr PAR SNPs}
\label{ss:pearson}

Consider the sex-combined genotype counts $(n_0,n_1,n_2)$ for an autosomal SNP (Table \ref{tab:notations}). Under the null of HWE, expected genotype frequencies are $\{(1-p)^2,\,2p(1-p),\,p^2\}$. When the allele frequency $p$ is unknown and estimated from the same sample, the classical Pearson HWE test is given by
\begin{equation}
\label{eq:pearson_auto_1df}
T_{\mbox{\footnotesize Pearson, Auto, }\hat{p}}= \frac{(n_0-n(1-\hat p)^2)^2}{n(1-\hat p)^2} + \frac{(n_1-n2\hat p(1-\hat p))^2}{n2\hat p(1-\hat p)} + \frac{(n_2-n\hat p^2)^2}{n\hat p^2} \: \overset{H_0}{\sim} \:  \chi^2_1,
\end{equation} 
where $\hat p = (n_1+2n_2)/2n$ is the sample estimate of $p$. The test $T_{\mbox{\footnotesize Pearson, Auto, }\hat{p}}$ has only 1 degree of freedom (df) because, in addition to being used for the HWE test, the same sample is used to estimate $p$, which leads to the loss of 1 df. 

When $p$ is externally specified, the same goodness-of-fit construction yields a 2 df test, 
\begin{equation} \label{eq:pearson_auto_2df}
T_{\mbox{\footnotesize Pearson, Auto, }p} =  \frac{(n_0 - n(1-p)^2)^2}{n(1-p)^2} + \frac{(n_1 - n2p(1-p))^2}{n2p(1-p)} + \frac{(n_2 - np^2)^2}{np^2}  \overset{H_0}{\sim} \chi^2_2.
\end{equation}

To clarify the distinction, we reformulate (\ref{eq:pearson_auto_2df}) as
\begin{equation} \label{eq:pearson_auto_2df_reform}
T_{\mbox{\footnotesize Pearson, Auto, }p} = \frac{\{\hat{p}_2 - \hat{p}^2 + (\hat{p} - p)^2\}^2}{\frac{1}{n}p^2(1-p)^2} + \frac{(\hat{p}-p)^2}{\frac{1}{2n} p(1-p)} \overset{H_0}{\sim} \chi^2_2, 
\end{equation}
and (\ref{eq:pearson_auto_1df}) can be rewritten as
\begin{equation} \label{eq:pearson_auto_1df_reform}
    T_{\mbox{\footnotesize Pearson, Auto, }\hat{p}} = \frac{(\hat{p}_2 - \hat{p}^2)^2 }{ \frac{1}{n} \hat{p}^2 (1-\hat{p} )^2 } \overset{H_0}{\sim} \chi^2_1.
\end{equation}
The key difference is that the 2 df test includes an additional component that evaluates the consistency between the sample allele frequency $\hat p$ and the externally specified value $p$, while the 1 df test focuses solely on the departure from HWE after estimating $p$ using the same data. See {Supplementary Notes 5} for details. 

Note that both Pearson-based tests (\ref{eq:pearson_auto_1df}) and (\ref{eq:pearson_auto_2df}) implicitly assume no sdMAF by using the same allele frequency for both sexes. This is typically reasonable for autosomes, but may not hold for Xchr PAR SNPs where sdMAF can occur \citep{wang_major_2022}. We return to PAR SNPs in Section \ref{s:par} and present Pearson-based tests that allow and do not allow sdMAF.

%%%%%%%%%%%%%%%%%%%%%%%%%%%%%%%%%%%%%%%%%%%%%%%%%%%%%%%%%%%%%%%%%%%%%%%%%%%%%%
%%%%%%%%%%%%%%%%%%%%%%%%%%%%%%%%%%%%%%%%%%%%%%%%%%%%%%%%%%%%%%%%%%%%%%%%%%%%%%
\subsection{Pearson $\chi^2$ tests of HWE for Xchr NPR SNPs} 
\label{sec:Pearson}

For Xchr NPR SNPs, the formulation of Pearson-based tests depends critically on two modeling choices: (i) whether HWE is defined through the structure of female genotypes (since males are hemizygous) and (ii) whether it is assumed that allele frequencies are shared between sexes (i.e., without sdMAF). These choices determine both the construction of the test statistic and its implicit null hypothesis. Consequently, different tests are not alternative implementations of a single procedure, but correspond to distinct inferential targets under different assumptions for Xchr NPR SNPs.

Table \ref{tab:pearson_tests} summarizes these Pearson-based tests, highlighting how differences in data usage and allele-frequency assumptions correspond to different null hypotheses. We detail each test in Sections \ref{sec:graffelman}--\ref{ss:hwewithsdmaf} below. 

\begin{landscape}
    \begin{table}[t]
\centering
\small
\newcolumntype{C}[1]{>{\centering\arraybackslash}p{#1}}
\newcolumntype{L}[1]{>{\raggedright\arraybackslash}p{#1}}
\begin{tabular}{C{2.7cm} L{0.2cm} C{3cm} L{3.4cm} || C{2.7cm} L{0.2cm} C{3.4cm} C{2.4cm}}
\hline \hline
Pearson Test & df & Data used in the goodness-of-fit component & Allele frequency used & RA test & df & $H_0$ tested & Model assumptions \\ \hline \hline 
$T_{\mathrm{Pearson,Auto},\hat p}$ (\ref{eq:pearson_auto_1df}) & 1 & Pooled autosomal genotype counts $(n_0,n_1,n_2)$ & $\hat p=(n_1+2n_2)/(2n)$ & $T_{\mathrm{RA,Auto},\hat p}$ (\ref{eq:raHWE_score}) & 1 & Testing HWE $H_0: \rho = 0$ & -- \\ \hline
 $T_{\mathrm{Pearson,Auto},p}$ (\ref{eq:pearson_auto_2df}) & 2 & Pooled autosomal genotype counts $(n_0,n_1,n_2)$ & fixed $p$ & $T_{\mathrm{RA,Auto}, p}$  & 2 & Testing HWE $H_0: \alpha = p \text{ and } \rho = 0$ & $\sigma^2 = \alpha(1-\alpha)$ \\ \hline
 $T_{\mathrm{Pearson,X\text{-}NPR_{F\&M}},\hat p}$ (\ref{eq:Pearson_xchr}) & 2 & Female $(f_0,f_1,f_2)$ and male $(m_0,m_2)$ counts & $\hat p=\dfrac{f_1+2f_2+m_2}{2f+m}$ & $T_{\mathrm{RA,X\text{-}NPR_{F\&M}},\hat p}$ (\ref{eq:raHWE_score_xchr}) & 2 & Testing HWE and no sdMAF $H_0: \gamma = 0 \text{ and } \rho = 0$ & --  \\ \hline
$T_{\mathrm{Pearson,X\text{-}NPR},\hat p}$ (\ref{eq:Pearson_?df}) & $\dagger$ & Female counts $(f_0,f_1,f_2)$ only & pooled $\hat p$ (uses males only via $\hat p$) & & & & \\ \hline
& & & & $T_{\mathrm{RA,X\text{-}NPR},\hat p}$ (\ref{eq:RA_no_sdMAF}) & 1 & Testing HWE $H_0: \rho = 0$ & No sdMAF $\gamma = 0$ \\ \hline
$T_{\mathrm{Pearson,X\text{-}NPR},\hat p_f}$ (\ref{eq:femaleonly}) & 1 & Female counts $(f_0,f_1,f_2)$ only & $\hat p_f=(f_1+2f_2)/(2f)$ & $T_{\mathrm{RA,X\text{-}NPR},\hat p_f}$ (\ref{eq:RA_sdMAF}) & 1 & Testing HWE  $H_0: \rho = 0$ & -- \\ \hline\hline
\end{tabular}

\caption{{\bf Unified summary of Pearson $\chi^2$-based and RA-based HWE tests under different assumptions regarding sdMAF and genotype structure.}
Here, the ``goodness-of-fit component'' indicates which observed genotype/allele counts enter the Pearson
$\chi^2$ sum. sdMAF denotes sex difference in allele frequency.
$\dagger$~$T_{\mathrm{Pearson,X\text{-}NPR},\hat p}$ has a weighted-sum-of-$\chi^2_1$ null (see text).}
\label{tab:pearson_tests}
\end{table}
\end{landscape}

%%%%%%%%%%%%%%%%%%%%%%%%%%%%%%%%%%%%%%%%%%%%%%%%%%%%%%%%%%%%%%%%%%%%%%%%%%%%%%
\subsubsection{Pearson $\chi^2_2$ test with both female and male samples using a sex-combined MAF estimate, $T_{\mbox{\footnotesize Pearson, X-NPR$_{F \& M}$, }\hat{p}}$}
\label{sec:graffelman}

\citet{graffelman2016testing} introduced a Pearson $\chi^2$-based tests for Xchr NPR SNPs, given in (\ref{eq:Pearson_xchr}), with a pooled allele-frequency estimate $\hat p$ defined in (\ref{eq:phat}). In $T_{\text{\footnotesize Pearson, X-NPR$_{F\&M}$, } \hat{p} }$, male data contribute both to the estimation of allele frequency $p$ and to a goodness-of-fit component that evaluates the assumption of no-sdMAF at the NPR SNP. Based on their empirical simulation studies, \citet{graffelman2016testing} reasoned that {\it ``rejection can occur if female genotype proportions deviate from HWE proportions, if female and male allele frequencies differ, or if both these phenomena occur simultaneously''}. 

\begin{equation}
\label{eq:Pearson_xchr}
\begin{aligned}
T_{\mbox{\footnotesize Pearson, X-NPR$_{F\&M}$, }\hat{p}} = & \frac{ (f_0 - f (1-\hat{p})^2)^2 }{f (1-\hat{p})^2} + \frac{ (f_1 - f 2 \hat{p}(1-\hat{p}))^2 }{f 2 \hat{p}(1-\hat{p})} + \frac{ (f_2 - f \hat{p}^2)^2 }{f \hat{p}^2} \\
+ & \frac{(m_0 - m (1-\hat{p}))^2}{m (1-\hat{p})} + \frac{(m_2 - m \hat{p})^2}{m \hat{p}} \overset{H_0}{\sim} \chi^2_2,
\end{aligned}
\end{equation} 
where $\hat{p}$ is the pooled sample allele frequency estimated using both sexes, 
\begin{equation} \label{eq:phat}
\hat{p} = \frac{f_1 + 2 f_2 + m_2 }{2f + m}. 
\end{equation}

Although not explicitly stated in \citet{graffelman2016testing}, the corresponding null hypothesis of $T_{\text{\footnotesize Pearson, X-NPR$_{F\&M}$, }\hat{p}}$, in fact, jointly assumes (i) HWE in females and (ii) no sdMAF ($p_f=p_m$). Consequently, rejection may arise from departures in either component. The precise relationship between this Pearson test statistic and its separate HWD and sdMAF components is established analytically in Section~\ref{subsec:link}.

%%%%%%%%%%%%%%%%%%%%%%%%%%%%%%%%%%%%%%%%%%%%%%%%%%%%%%%%%%%%%%%%%%%%%%%%%%%%%%
\subsubsection{Pearson $\chi^2$-based test with female samples only but using a sex-combined MAF estimate (i.e., assuming no sdMAF), $T_{\mbox{\footnotesize Pearson, X-NPR, }\hat{p}}$}
\label{ss:nosdmaf}

If the goal is to test whether female genotype counts conform to Hardy--Weinberg proportions {\it under the assumption of no sdMAF}, then only female genotype counts enter the goodness-of-fit component, and male data contribute only through the pooled allele-frequency estimate. The null hypothesis is therefore HWE in females, assuming no sdMAF. The corresponding Pearson $\chi^2$ statistic is $T_{\mathrm{Pearson,X\text{-}NPR},\hat p}$ in (\ref{eq:Pearson_?df}) with $\hat p$ defined by (\ref{eq:phat}). 

\begin{equation} \label{eq:Pearson_?df}
\begin{aligned}
T_{\mbox{\footnotesize Pearson, X-NPR, }\hat{p}} & = \frac{ (f_0 - f (1-\hat{p})^2)^2 }{f (1-\hat{p})^2} + \frac{ (f_1 - f 2 \hat{p}(1-\hat{p}))^2 }{f 2 \hat{p}(1-\hat{p})} + \frac{ (f_2 - f \hat{p}^2)^2 }{f \hat{p}^2}.
\end{aligned}
\end{equation}

\noindent \textbf{Remark (effective degrees of freedom).} 
Unlike the classical autosomal Pearson test, the null distribution of 
$T_{\text{\footnotesize Pearson, X-NPR, }\hat{p}}$
is not a standard $\chi^2_k$ distribution because the pooled allele-frequency estimate contains both internal female and external male information. In Section \ref{s:methods}, we show that it can be expressed as a weighted sum of two independent $\chi^2_1$ variables.

%%%%%%%%%%%%%%%%%%%%%%%%%%%%%%%%%%%%%%%%%%%%%%%%%%%%%%%%%%%%%%%%%%%%%%%%%%%%%%
\subsubsection{A Pearson $\chi^2_1$ test using female samples only (allowing for sdMAF), $T_{\mbox{\footnotesize Pearson, X-NPR, }\hat{p}_f}$}
\label{ss:hwewithsdmaf}

When sdMAF is allowed for the Xchr NPR SNP of interest, male data do not contribute to HWE testing because HWE is defined through the structure of diploid genotypes, which is only available in females in Xchr NPR.
\begin{equation}
T_{\mbox{\footnotesize Pearson, X-NPR, }\hat{p}_f} =  \frac{ (f_0 - f (1-\hat{p}_f)^2)^2 }{f (1-\hat{p}_f)^2} + \frac{ (f_1 - f 2 \hat{p}_f(1-\hat{p}_f))^2 }{f 2 \hat{p}_f(1-\hat{p}_f)} + \frac{ (f_2 - f \hat{p}_f^2)^2 }{f \hat{p}_f^2} \overset{H_0}{\sim} \chi^2_1.
\label{eq:femaleonly}
\end{equation}
This female-only approach is widely used in practice regardless of sdMAF status (e.g., PLINK \citep{purcell2007plink}).

The Pearson $\chi^2$-based formulations presented in Section~\ref{s:overview} demonstrate that multiple valid HWE testing procedures arise for Xchr SNPs under different modeling assumptions regarding genotype structures and allele frequencies. In particular, for Xchr NPR SNPs, distinct choices regarding how male data are incorporated and whether allele frequencies are assumed to be shared between sexes yield different test statistics with different inferential targets.

Although these formulations are all grounded in the classical goodness-of-fit framework, their relationships are not immediately transparent, and their underlying assumptions are often not explicit. This can make it difficult to interpret the differences between tests or extend them to more complex settings involving covariates or heterogeneous populations.

To address these issues, we introduce a unified regression-based framework for HWE inference on the X chromosome in the next section. By formulating HWE testing as an assessment of allele-level dependence, this approach provides a common parameterization of HWD, clarifies the assumptions underlying existing tests, and establishes direct connections among them. This framework also enables greater flexibility, including the incorporation of sdMAF and covariate effects within a single coherent modeling structure.

%%%%%%%%%%%%%%%%%%%%%%%%%%%%%%%%%%%%%%%%%%%%%%%%%%%%%%%%%%%%%%%%%%%%%%%%%%%%%%
%%%%%%%%%%%%%%%%%%%%%%%%%%%%%%%%%%%%%%%%%%%%%%%%%%%%%%%%%%%%%%%%%%%%%%%%%%%%%%
%%%%%%%%%%%%%%%%%%%%%%%%%%%%%%%%%%%%%%%%%%%%%%%%%%%%%%%%%%%%%%%%%%%%%%%%%%%%%%
\section{A regression-based score test of HWE for the X chromosome}
\label{s:methods}

To provide a unified perspective on HWE testing for X-chromosomal SNPs, we formulate HWE inference within a regression-based framework based on the robust allele-based (RA) model. Rather than treating HWE testing as a goodness-of-fit problem defined by genotype counts, this approach operates at the allele level and characterizes HWE as a form of independence between paired alleles within individuals. This formulation directly parameterizes HWD, clarifies the assumptions underlying the existing Pearson $\chi^2$ tests, and provides a unified foundation for Xchr HWE inference under sdMAF.

\citet{zhang2022generalized} proposed the RA regression framework for allelic association testing in complex data settings for autosomes. The framework represents each individual by two alleles and models their joint distribution, allowing for a direct characterization of dependence between alleles within an individual. In this formulation, the HWE corresponds to the independence between the two alleles in a genotype, while the HWD is captured by deviation from independence. The RA framework, therefore, provides a natural regression-based approach to HWE testing.
%%%%%%%%%%%%%%%%%%%%%%%%%%%%%%%%%%%%%%%%%%%%%%%%%%%%%%%%%%%%%%%%%%%%%%%%%%%%%%
%%%%%%%%%%%%%%%%%%%%%%%%%%%%%%%%%%%%%%%%%%%%%%%%%%%%%%%%%%%%%%%%%%%%%%%%%%%%%%
\subsection{A regression-based score test of HWE for autosomal SNPs}
\label{subsec:RA}

We represent each individual by two alleles at a given SNP and model their joint distribution. Let $(G_{i1}, G_{i2})$ denote the two alleles carried by individual $i$, where each allele takes value 0 or 1, corresponding to alleles $a$ and $A$, respectively. 

For a bi-allelic autosomal SNP, the RA framework represents the unordered genotype as an ordered allele pair
\begin{equation}
\label{eq:hwe_gi12}
(G_{i1}, G_{i2}) = 
\begin{cases}
(0, 0) \hspace{0.3in} \text{if genotype is $aa$ } \\
(0, 1) \hspace{0.3in} \text{if genotype is $Aa$ and $c_i=0$} \\
(1, 0) \hspace{0.3in} \text{if genotype is $Aa$ and $c_i=1$} \\
(1, 1) \hspace{0.3in} \text{if genotype is $AA$, } \\
\end{cases}
\end{equation}
where $c_i = 1$ if the flip of a fair coin is head and $c_i = 0$ otherwise. The flip of a fair coin ensures that the phased genotype $Aa$ and $aA$ occur equally likely when an unordered heterozygous genotype $Aa$ is observed.

For an independent sample with no phenotypes, the RA regression framework is defined as 
\begin{equation} \label{eq:ra_full}
\begin{pmatrix}
G_{i1} \\
G_{i2} \\
\end{pmatrix}
  =(\alpha + \gamma Z_i)  \begin{pmatrix}
1 \\
1 \\
\end{pmatrix}  + \begin{pmatrix}
\epsilon_{i1} \\
\epsilon_{i2} \\
\end{pmatrix}, 
\hspace{0.3cm} \text{where} \hspace{0.15cm}
\begin{pmatrix}
\epsilon_{i1} \\
\epsilon_{i2} \\
\end{pmatrix}
 \overset{iid}{\sim} N(0, \sigma^2 
\begin{pmatrix}
1 & \rho \\
\rho & 1 \\
\end{pmatrix}),
\end{equation}
where $Z$ is the covariate, and $\rho$ in the variance-covariance matrix explicitly models the potential HWD in the sample. Testing whether HWE holds in the sample reduces to testing $H_0: \rho = 0$. Although the outcome is binary \citep{chen1983score}, Gaussian working models are well established in allele-based inference \citep{zhang2022generalized, zhang2023leveraging, zhang2022linear, wang2024population}. 

This formulation makes explicit that HWE is fundamentally a statement about within-individual allele independence, rather than a property of genotype counts. If we further assume $\gamma = 0$ in (\ref{eq:ra_full}), testing $H_0: \rho = 0$ reduces to testing HWE in an independent sample without adjusting for covariate effects, and the score test statistic is 
\begin{equation} \label{eq:raHWE_score}
T_{\mbox{\footnotesize RA, Auto, }\hat{p}} = n \hat{\rho}^2 = \frac{\hat{\delta}^2}{\frac{1}{n} \hat{p}^2(1-\hat{p})^2}  = \frac{(\hat{p}_2 - \hat{p}^2)^2}{\frac{1}{n} \hat{p}^2(1-\hat{p})^2}  \overset{H_0}{\sim} \chi^2_1, 
\end{equation}
where $\hat{\delta} = (\hat{p}_2 - \hat{p}^2)$ is a sample estimate of the classical measure of HWD \citep{weir1996genetic}, and $\hat{\rho} = \hat{\delta}/(\hat{p}(1-\hat{p}))$ estimates the HWD magnitude scaled by the sample allele frequency. 

\citet{zhang2022generalized} showed analytically that $T_{\mbox{\footnotesize RA, Auto, }\hat{p}}$ in (\ref{eq:raHWE_score}) is identical to the classical Pearson $\chi^2_1$ test of HWE for bi-allelic autosomal SNPs in a finite sample. This result provides an alternative interpretation of the Pearson test as a score test for allele-level dependence in the RA framework. It also reveals that HWE testing fundamentally corresponds to assessing whether $\rho=0$. 

%%%%%%%%%%%%%%%%%%%%%%%%%%%%%%%%%%%%%%%%%%%%%%%%%%%%%%%%%%%%%%%%%%%%%%%%%%%%%%
%%%%%%%%%%%%%%%%%%%%%%%%%%%%%%%%%%%%%%%%%%%%%%%%%%%%%%%%%%%%%%%%%%%%%%%%%%%%%% 
\subsection{The RA regression model for Xchr NPR SNPs}

We first extend the RA framework to X-chromosomal SNPs in the NPRs, where genotype structures differ between sexes. Females are diploid and contribute two alleles per individual, while males are hemizygous and contribute one allele. This distinction requires modifying the regression formulation to accommodate sex-specific contributions.

Let $G$ denote the genotype and represent each individual by allele indicators under the RA framework. For a bi-allelic Xchr NPR SNP, the genotype of a female sample is partitioned into two alleles $(G_{fi1}, G_{fi2})$ in the same manner as for autosomal SNPs in (\ref{eq:hwe_gi12}). For a male hemizygous genotype, we define the allele indicator $G_{mj} = 1$ if the genotype is $A$, and $0$ otherwise. Under HWE, the two alleles in a female are independent, whereas no within-individual dependence can be defined for males.

The general RA framework for Xchr NPR SNPs is
\begin{equation} \label{eq:RA_NPR_general}
    \begin{pmatrix} G_{fi1} \\ G_{fi2} \\ G_{mj} \end{pmatrix} = \alpha \begin{pmatrix} 1 \\ 1 \\ 1 \\ \end{pmatrix}  + \gamma \begin{pmatrix}
\mbox{Sex}_i \\
\mbox{Sex}_i \\
\mbox{Sex}_j \\
\end{pmatrix} + \begin{pmatrix}
\epsilon_{i1} \\
\epsilon_{i2} \\
\epsilon_{j}  \\
\end{pmatrix}, \: \: \mbox{where} \: \: \begin{pmatrix}
\epsilon_{i1} \\
\epsilon_{i2} \\
\epsilon_{j}  \\
\end{pmatrix} \sim N(\bzero,  \begin{pmatrix} \sigma_f^2 & \rho_f \sigma_f^2 & 0 \\ \rho_f \sigma_f^2 & \sigma_f^2 & 0 \\ 
0 & 0 & \sigma_m^2 \\
\end{pmatrix}),
\end{equation}
where $\text{Sex} = 1$ for females and $0$ for males; $\sigma_f^2$ and $\sigma_m^2$ are the sex-specific variances for females and males, respectively; and $\rho_f$ measures HWD in the female population.

\noindent\textbf{Remark.}
The roles of the female and male samples are determined directly by the model structure, rather than being imposed through the test construction, in contrast to Pearson-based formulations.

Let $p_f$ and $p_m$ denote the allele frequencies in females and males. Unlike the autosomal case, allele frequencies may differ between sexes on the X chromosome. If the goal is to test HWE in females, the RA framework can be applied directly to female genotypes by testing $H_0: \rho_f = 0$. When male data are incorporated, their role depends on whether one assumes $p_f = p_m$. Under the assumption of no sdMAF, male data contribute to the estimation of a pooled allele frequency. Otherwise, male and female allele frequencies must be modeled separately.

The HWE testing goal and the sdMAF assumptions for an Xchr NPR SNP can be directly linked to the parameters in (\ref{eq:RA_NPR_general}), resulting in explicit null hypotheses. The inclusion of female and male samples in both the testing and allele frequency estimation is naturally handled within the regression framework, which also ensures well-defined null distributions for the resulting test statistics.

In the following sections, we introduce three variants of the RA framework for Xchr NPR SNPs corresponding to different testing goals and assumptions: (i) jointly testing HWE and sdMAF; (ii) testing HWE under the assumption of no sdMAF; and (iii) testing HWE allowing sdMAF; see Table \ref{tab:pearson_tests} for a summary of the test statistics. We also establish their connections to the Pearson $\chi^2$-based tests described in Section \ref{sec:Pearson}.

%%%%%%%%%%%%%%%%%%%%%%%%%%%%%%%%%%%%%%%%%%%%%%%%%%%%%%%%%%%%%%%%%%%%%%%%%%%%%%
%%%%%%%%%%%%%%%%%%%%%%%%%%%%%%%%%%%%%%%%%%%%%%%%%%%%%%%%%%%%%%%%%%%%%%%%%%%%%%
\subsection{Jointly testing HWE and sdMAF} \label{subsec:2df_tests}

If the goal is to jointly test HWE and sdMAF at an Xchr NPR SNP, the null hypothesis is, 
\begin{equation} \label{eq:null_2df}
    H_0: \gamma = 0 \: \: \text{and} \: \: \rho_f = 0. 
\end{equation}
Under this null, the general RA model in (\ref{eq:RA_NPR_general}) is reduced by assuming $\sigma_f^2 = \sigma_m^2 = \sigma^2$, resulting in
\begin{equation} \label{eq:RA_2df}
\begin{pmatrix} G_{fi1} \\ G_{fi2} \\ G_{mj} \\ \end{pmatrix} = \alpha \begin{pmatrix} 1 \\ 1 \\ 1 \\ \end{pmatrix} + \gamma \begin{pmatrix} \mbox{Sex}_i \\ \mbox{Sex}_i \\ \mbox{Sex}_j \\ \end{pmatrix} + 
    \begin{pmatrix} \epsilon_{i1} \\ \epsilon_{i2} \\ \epsilon_{j} \\ \end{pmatrix}, \: \: \mbox{where} \: \: \begin{pmatrix} \epsilon_{i1} \\ \epsilon_{i2} \\ \epsilon_{j} \\ \end{pmatrix} \sim N(\bzero, \sigma^2 \begin{pmatrix} 1 & \rho_f & 0 \\ \rho_f & 1 & 0 \\ 0 & 0 & 1 \\ \end{pmatrix}),
\end{equation}
The score test for (\ref{eq:null_2df}) yields
\begin{equation} \label{eq:raHWE_score_xchr}
\begin{aligned}
T_{\mbox{\footnotesize RA, X-NPR$_{F\&M}$, }\hat{p}} 
& = \frac{\{ \hat{\delta}_f + (\frac{m}{2f + m})^2 ( \hat{p}_f - \hat{p}_m)^2 \}^2 }{\frac{1}{f}\hat{p}^2(1-\hat{p})^2} + \frac{ (\hat{p}_f - \hat{p}_m)^2}{(\frac{1}{2 f} + \frac{1}{m})\hat{p}(1-\hat{p})} \overset{H_0}{\sim} \chi^2_2, 
\end{aligned}
\end{equation}
where $\hat{p} = (2f_2 + f_1 + m_2)/(2f + m)$, $\hat{p}_f = (2f_2 + f_1)/(2f)$ and $\hat{p}_m = m_2/m$ as in Table \ref{tab:notations}, $\hat{\delta}_f = \hat{p}_2 - \hat{p}_f^2$, where $\hat{p}_2 = f_2/f$.

This formulation jointly evaluates HWE and sdMAF under a common regression framework. Its analytical decomposition and relationship to the Pearson-based formulation are developed in Section \ref{subsec:link}.

%%%%%%%%%%%%%%%%%%%%%%%%%%%%%%%%%%%%%%%%%%%%%%%%%%%%%%%%%%%%%%%%%%%%%%%%%%%%%%
%%%%%%%%%%%%%%%%%%%%%%%%%%%%%%%%%%%%%%%%%%%%%%%%%%%%%%%%%%%%%%%%%%%%%%%%%%%%%%
\subsection{Testing HWE under the assumption of no sdMAF}
\label{subsec:1df_no_sdMAF}

If the goal is to test HWE under the assumption that there is no sdMAF at an Xchr NPR SNP, then the model imposes $p_f=p_m$. Equivalently, the sex effect is removed from the general RA model in (\ref{eq:RA_NPR_general}), yielding a reduced model with a pooled allele frequency and a common variance structure:
\begin{equation} \label{eq:modelwithnosdmaf}
\begin{pmatrix} G_{fi1} \\ G_{fi2} \\ G_{mj} \\ \end{pmatrix} = \alpha \begin{pmatrix} 1 \\ 1 \\ 1 \\ \end{pmatrix} + \begin{pmatrix} \epsilon_{i1} \\ \epsilon_{i2} \\ \epsilon_{j} \\ \end{pmatrix}, \: \: \mbox{where} \: \: \begin{pmatrix} \epsilon_{i1} \\ \epsilon_{i2} \\ \epsilon_{j} \\ \end{pmatrix} \sim N(0, \sigma^2 \begin{pmatrix} 1 & \rho_f & 0 \\ \rho_f & 1 & 0 \\ 0 & 0 & 1 \\ \end{pmatrix}). 
\end{equation}

Under this reduced model, testing HWE reduces to testing
\begin{equation} \label{eq:nullwithnosdmaf}
H_0: \rho_f = 0.
\end{equation}

The corresponding score test is
\begin{equation} \label{eq:RA_no_sdMAF}
    T_{\mbox{\footnotesize RA, X-NPR, }\hat{p}} =
\frac{\{\hat{\delta}_f + ( \hat{p} - \hat{p}_f)^2 \}^2 }{\frac{1}{f} \hat{p}^2(1-\hat{p})^2} = \frac{\{\hat{\delta}_f + (\frac{m}{2f + m})^2 (\hat{p}_f - \hat{p}_m)^2 \}^2 }{\frac{1}{f} \hat{p}^2 (1-\hat{p})^2} \overset{H_0}{\sim} \chi^2_1, 
\end{equation}
which imposes the no-sdMAF assumption and evaluates HWE using a single allele frequency shared between sexes, in contrast to the joint test in Section \ref{subsec:2df_tests}. Thus, test (\ref{eq:RA_no_sdMAF}) focuses on the departures from HWE with the constraint of  $p_f=p_m$. 

This formulation is related to the Pearson $\chi^2$-based test described in Section \ref{ss:nosdmaf}, where both female and male samples are used to estimate the pooled allele frequency, but only the diploid female genotypes contribute to the HWE component of the test.

%%%%%%%%%%%%%%%%%%%%%%%%%%%%%%%%%%%%%%%%%%%%%%%%%%%%%%%%%%%%%%%%%%%%%%%%%%%%%%
%%%%%%%%%%%%%%%%%%%%%%%%%%%%%%%%%%%%%%%%%%%%%%%%%%%%%%%%%%%%%%%%%%%%%%%%%%%%%%
\subsection{Testing HWE while allowing for sdMAF}
\label{subsec:1df_sdMAF}

If sdMAF is allowed for the Xchr NPR SNP of interest, we retain the general RA framework in (\ref{eq:RA_NPR_general}), where allele frequencies can differ between sexes. In this setting, $\alpha$, $\gamma$, $\sigma_f^2$, and $\sigma_m^2$ are treated as nuisance parameters.

Testing HWE then reduces to testing 
\begin{equation} \label{eq:nullwithsdmaf}
H_0: \rho_f = 0.
\end{equation}

The corresponding score test statistic is
\begin{equation} \label{eq:RA_sdMAF}
T_{\mbox{\footnotesize RA, X-NPR, }\hat{p}_f} = \frac{\hat{\delta}_f^2}{\frac{1}{f} \hat{p}_f^2 (1-\hat{p}_f)^2} \overset{H_0}{\sim} \chi^2_1,
\end{equation}
which depends only on the female data.

This test is equivalent to applying the classical autosomal HWE test to female samples only. In contrast to Section \ref{subsec:1df_no_sdMAF}, male data do not contribute to allele frequency estimation or the HWE test statistic, because the presence of sdMAF precludes sex-pooled allele frequency estimate, and the hemizygous male genotypes make HWE not meaningful for males. This is consistent with the Pearson $\chi^2$-based test described in Section \ref{ss:hwewithsdmaf} and highlights a key feature of the RA framework: the role of male samples is determined directly by the model assumptions, without requiring ad hoc adjustments.

%%%%%%%%%%%%%%%%%%%%%%%%%%%%%%%%%%%%%%%%%%%%%%%%%%%%%%%%%%%%%%%%%%%%%%%%%%%%%%
%%%%%%%%%%%%%%%%%%%%%%%%%%%%%%%%%%%%%%%%%%%%%%%%%%%%%%%%%%%%%%%%%%%%%%%%%%%%%%
\subsection{Unified interpretation of RA-based and Pearson $\chi^2$ tests for Xchr NPR SNPs}
\label{subsec:link}

We have introduced three RA-based tests in Sections \ref{subsec:2df_tests}--\ref{subsec:1df_sdMAF}, and three Pearson $\chi^2$ tests in Section \ref{sec:Pearson} for Xchr NPR SNPs under different assumptions on HWE and sdMAF. In this section, we establish their analytical connections and clarify their relationships in a unified regression framework.

\subsubsection{Analytical connections between RA-based and Pearson $\chi^2$ tests} \label{subsubsec:equivalence}
The RA-based tests in Sections \ref{subsec:2df_tests}--\ref{subsec:1df_sdMAF} correspond to the Pearson $\chi^2$ tests in Section \ref{sec:Pearson} in three distinct settings, with each test arising from specific modeling assumptions about HWE and sdMAF. This establishes a unified regression-based interpretation of the classical tests.

First, the RA-based test that allows sdMAF, $T_{\mbox{\footnotesize RA, X-NPR, } \hat{p}_f}$,
is identical to the Pearson test $T_{\mbox{\footnotesize Pearson, X-NPR, } \hat{p}_f}$ in Section \ref{ss:hwewithsdmaf}, 
\begin{equation*}
    T_{\mbox{\footnotesize RA, X-NPR, }\hat{p}_f} = T_{\mbox{\footnotesize Pearson, X-NPR, }\hat{p}_f}, 
\end{equation*}
as both reduce to the classical autosomal HWE test applied to female samples only; see Section \ref{subsec:RA} and {Supplementary Notes 2.4}.

Second, the joint RA-based test, $T_{\mbox{\footnotesize RA, X-NPR$_{F\&M}$, } \hat{p}}$, is identical to the Pearson test $T_{\mbox{\footnotesize Pearson, X-NPR$_{F\&M}$, } \hat{p}}$ in Section \ref{sec:graffelman},
\begin{equation*}
    T_{\mbox{\footnotesize RA, X-NPR$_{F\&M}$, }\hat{p}} = T_{\mbox{\footnotesize Pearson, X-NPR$_{F\&M}$, }\hat{p}}. 
\end{equation*}
This equivalence shows that the Pearson test is a joint test of HWE and sdMAF, as its statistic decomposes into components corresponding to HWD and sdMAF ({Supplementary Notes 2.2}), providing a formal statistical justification for the empirical observations of \citet{graffelman2016testing}.

Third, the RA-based test under the assumption of no sdMAF, $T_{\mbox{\footnotesize RA, X-NPR, } \hat{p}}$, is closely related to the Pearson test $T_{\mbox{\footnotesize Pearson, X-NPR, } \hat{p}}$ in Section \ref{ss:nosdmaf}, although they are not identical. As shown in Section \ref{sec:linkRA}, the Pearson statistic $T_{\mbox{\footnotesize Pearson, X-NPR, } \hat{p}}$ can be expressed as a weighted sum of the RA-based HWE test and an sdMAF test component, yielding a non-standard null distribution that depends on the sample composition.

%%%%%%%%%%%%%%%%%%%%%%%%%%%%%%%%%%%%%%%%%%%%%%%%%%%%%%%%%%%%%%%%%%%%%%%%%%%%%%
\subsubsection{Decomposition and relationships among tests} \label{sec:linkRA}

The three RA-based tests are intrinsically related through the underlying regression model in (\ref{eq:RA_NPR_general}). The joint test in Section \ref{subsec:2df_tests} is based on the full model and jointly assesses both HWD and sdMAF. The test in Section \ref{subsec:1df_no_sdMAF} is obtained by imposing the constraint $p_f = p_m$, thus removing the sex effect and reducing the model to a pooled-allele-frequency model. In contrast, the test in Section \ref{subsec:1df_sdMAF} treats sex-specific allele frequencies as nuisance parameters and focuses solely on HWD in females.

These relationships show that differences in degrees of freedom and the role of male samples arise directly from the model assumptions, rather than from ad hoc modifications to the classical HWE tests. Their analytical decomposition further clarifies these relationships.

The joint RA-based test admits the decomposition
\begin{equation} \label{eq:link_joint}
\begin{aligned}
    T_{\mbox{\footnotesize RA, X-NPR}_{F\&M}, \: \hat{p}} & = \frac{\{ \hat{\delta}_f + (\frac{m}{2f + m})^2 ( \hat{p}_f - \hat{p}_m)^2 \}^2 }{\frac{1}{f}\hat{p}^2(1-\hat{p})^2} + \frac{ (\hat{p}_f - \hat{p}_m)^2}{(\frac{1}{2 f} + \frac{1}{m})\hat{p}(1-\hat{p})} \\
    & = T_{\mbox{\footnotesize RA, X-NPR}, \: \hat{p}} + T_{\text{\footnotesize sdMAF, X-NPR, HWE}},
\end{aligned}
\end{equation}
where $T_{\text{sdMAF, X-NPR, HWE}}$ is the test of sdMAF under HWE. Under the joint null hypothesis of HWE and no sdMAF, these two components are independent of each other and each follows a standard $\chi^2_1$ distribution; see {Supplementary Notes 1.4}. This decomposition shows explicitly that the Pearson test (in Section \ref{sec:graffelman}) simultaneously evaluates HWD and sdMAF within a unified regression framework.

Similarly, the Pearson test, under the assumption of no sdMAF in Section \ref{ss:nosdmaf}, can be expressed as
\begin{equation} \label{eq:link_pearson_no_sdmaf}
\begin{aligned}
    T_{\mbox{\footnotesize Pearson, X-NPR, }\hat{p}} & = \frac{\{ \hat{\delta}_f + (\frac{m}{2f + m})^2 ( \hat{p}_f - \hat{p}_m)^2 \}^2 }{\frac{1}{f}\hat{p}^2(1-\hat{p})^2} + \frac{(\hat{p}_f - \hat{p})^2}{\frac{1}{2f} \hat{p}(1-\hat{p})} \\
    & = T_{\text{\footnotesize RA, X-NPR, }\hat{p}} + \frac{m}{2f+m} T_{\text{\footnotesize sdMAF, X-NPR, HWE}},
\end{aligned}
\end{equation}
which reveals that it is a weighted sum of two independent $\chi^2_1$ distributions, with weight $\frac{m}{2f+m}$, where $m$ is the number of males and $f$ is the number of females; see {Supplementary Notes 2.3}. This explains why its null distribution is not a standard $\chi^2_k$ distribution and depends on the sample composition.

%%%%%%%%%%%%%%%%%%%%%%%%%%%%%%%%%%%%%%%%%%%%%%%%%%%%%%%%%%%%%%%%%%%%%%%%%%%%%%
\subsubsection{Implications for Xchr HWE inference} 

Although both $T_{\mbox{\footnotesize RA, X-NPR, } \hat{p}}$ and $T_{\mbox{\footnotesize Pearson, X-NPR, } \hat{p}}$ are intended to test HWE under the assumption that there is no sdMAF, they differ fundamentally in construction. The Pearson test implicitly incorporates an sdMAF component via the pooled allele frequency, resulting in a sample-size-dependent null distribution. In contrast, the RA-based test directly specifies the null hypothesis via model parameters, resulting in a well-defined, asymptotic $\chi^2_1$ distribution under the null hypothesis.

Taken together, these results show that the seemingly distinct Pearson-based tests in Section \ref{sec:Pearson} are unified within the RA framework as score tests derived from a common model under different assumptions. More broadly, Xchr HWE tests form a structured family of inferential procedures whose null hypotheses, degrees of freedom, and data usage are determined directly by explicit modeling assumptions regarding HWE and sdMAF. This unified perspective clarifies existing methods, explains their differing behaviours, and establishes a general statistical foundation for HWE inference on the X chromosome.

%%%%%%%%%%%%%%%%%%%%%%%%%%%%%%%%%%%%%%%%%%%%%%%%%%%%%%%%%%%%%%%%%%%%%%%%%%%%%%
%%%%%%%%%%%%%%%%%%%%%%%%%%%%%%%%%%%%%%%%%%%%%%%%%%%%%%%%%%%%%%%%%%%%%%%%%%%%%%
%%%%%%%%%%%%%%%%%%%%%%%%%%%%%%%%%%%%%%%%%%%%%%%%%%%%%%%%%%%%%%%%%%%%%%%%%%%%%%
\section{Tests of HWE for Xchr PAR SNPs} \label{s:par}

In the PARs of the X chromosome, both females and males are diploids, and the genotype structures resemble that of autosomal SNPs. However, despite this apparent similarity, the HWE inference in the Xchr PARs can be complicated by sdMAF, particularly near the PAR–NPR boundaries where sdMAF has been empirically observed \citep{wang_major_2022,chen2023comprehensive}. Consequently, the classical HWE test using sex-pooled data for autosomal SNPs are valid only under additional assumptions, most notably the absence of sdMAF. The RA framework naturally extends to PAR SNPs by explicitly accommodating these assumptions while preserving a unified inferential structure across Xchr regions.

To accommodate potential sdMAF at Xchr PAR SNPs, we formulate HWE testing within the general RA regression framework. For a bi-allelic PAR SNP, both females and males contribute two alleles per individual, with $(G_{fi1}, G_{fi2})^T$ denoting the alleles for female sample $i$ and $(G_{mj1}, G_{mj2})^T$ for male sample $j$, both partitioned as in (\ref{eq:hwe_gi12}). 

The RA model can be written as
\begin{equation} \label{eq:par_sdmaf}
    \begin{pmatrix}
        G_{fi1} \\
        G_{fi2} \\
        G_{mj1} \\
        G_{mj2} \\
    \end{pmatrix} = 
\alpha  \begin{pmatrix}
        1 \\
        1 \\
        1 \\
        1 \\
    \end{pmatrix} + 
 \gamma   \begin{pmatrix}
    \text{Sex}_{i} \\
    \text{Sex}_{i} \\
    \text{Sex}_{j} \\
    \text{Sex}_{j} \\
    \end{pmatrix} + 
    \begin{pmatrix}
        \epsilon_{fi1} \\
        \epsilon_{fi2} \\
        \epsilon_{mj1} \\
        \epsilon_{mj2} \\
    \end{pmatrix},
\end{equation}
where $\text{Sex}=1$ for females and $0$ for males, and the error terms follow
\begin{equation}
    \begin{pmatrix}
        \epsilon_{fi1} \\
        \epsilon_{fi2} \\
        \epsilon_{mj1} \\
        \epsilon_{mj2} \\
    \end{pmatrix} \sim 
    N(\bzero, \begin{pmatrix}
        \sigma^2_f & \sigma^2_f \rho_f & 0 & 0 \\
          \sigma^2_f \rho_f &  \sigma^2_f  & 0 & 0 \\
        0 & 0  &   \sigma^2_m & \sigma^2_m \rho_m \\
    0 & 0  & \sigma^2_m \rho_m &   \sigma^2_m \\
    \end{pmatrix}.
\end{equation}
Here, $\rho_f$ and $\rho_m$ quantify the Hardy--Weinberg disequilibrium in females and males, respectively. Allowing for sdMAF corresponds to $\gamma \neq 0$, permitting allele frequencies to differ between sexes. Thus, as in the NPR setting, HWE testing in PAR regions can be formulated as a structured score-testing problem under explicit assumptions regarding sdMAF, with different testing procedures corresponding directly to different model constraints.

To test HWE at a PAR SNP while allowing for sdMAF, we evaluate
\begin{equation} \label{eq:null_par_sdmaf}
   H_0: \rho_f = \rho_m = 0,
\end{equation}
and the corresponding score test statistic is 
\begin{equation}
    T_{\text{\footnotesize RA, X-PAR, }\hat{p}_f, \: \hat{p}_m}  = \frac{\hat{\delta}_f^2}{\frac{1}{f} \hat{p}_f^2(1-\hat{p}_f)^2} + \frac{\hat{\delta}_m^2}{\frac{1}{m} \hat{p}_m^2(1-\hat{p}_m)^2} = f \hat{\rho}_f^2 + m \hat{\rho}_m^2 \overset{H_0}{\sim} \chi^2_2,
\end{equation}
where $\hat{p}_f = (2f_2 + f_1)/(2f)$ and $\hat{p}_m = (2m_2 + m_1)/(2m)$ are the sample allele frequencies in females and males, respectively, as in Table \ref{tab:notations}; $\hat{\delta}_f = f_2/f - \hat{p}_f^2$ and $\hat{\delta}_m = m_2/m - \hat{p}_m^2$ are the sex-specific HWD estimates. 

This test can be interpreted as the sum of two independent sex-specific autosomal HWE tests, one for females and one for males, thereby generalizing the autosomal framework of \cite{troendle1994note}. More broadly, it clarifies how the RA framework accommodates distinct biological and genomic structures through transparent model specification rather than ad hoc test modification.

If we further assume that there is no sdMAF at the PAR SNP of interest, the general RA model in (\ref{eq:par_sdmaf})  simplifies by imposing the same allele frequency and variance structure between sexes, where $\gamma = 0$, $\sigma_f^2 = \sigma_m^2 = \sigma^2$, and $\rho_f = \rho_m = \rho$. The resulting test reduces to a sex-pooled HWE test analogous to the classical autosomal Pearson $\chi^2_1$ test; see {Supplementary Notes 3.1.2 and 3.2.2--3.2.3} for details. Thus, the autosomal formulation emerges as a special case of the broader RA framework under explicit no-sdMAF assumption. This extension to Xchr PAR regions further demonstrates that the proposed general framework provides a unified statistical foundation for HWE inference across the whole genome.

%%%%%%%%%%%%%%%%%%%%%%%%%%%%%%%%%%%%%%%%%%%%%%%%%%%%%%%%%%%%%%%%%%%%%%%%%%%%%%
%%%%%%%%%%%%%%%%%%%%%%%%%%%%%%%%%%%%%%%%%%%%%%%%%%%%%%%%%%%%%%%%%%%%%%%%%%%%%%
%%%%%%%%%%%%%%%%%%%%%%%%%%%%%%%%%%%%%%%%%%%%%%%%%%%%%%%%%%%%%%%%%%%%%%%%%%%%%%
\section{Simulation studies and real-data applications}
\label{s:emprical}

%%%%%%%%%%%%%%%%%%%%%%%%%%%%%%%%%%%%%%%%%%%%%%%%%%%%%%%%%%%%%%%%%%%%%%%%%%%%%%
%%%%%%%%%%%%%%%%%%%%%%%%%%%%%%%%%%%%%%%%%%%%%%%%%%%%%%%%%%%%%%%%%%%%%%%%%%%%%%
\subsection{Simulation setup for Xchr NPR SNPs}
We conducted simulation studies to evaluate the type I error (T1E) and power of the proposed RA-based HWE tests for Xchr NPR SNPs under a range of HWD and sdMAF scenarios. Our primary focus was on Xchr NPR SNPs, where sex-specific genotype structures create the greatest inferential complexity; corresponding results for Xchr PAR SNPs are presented in {Supplementary Table S5} and {Supplementary Figures S9--S11}.

We compared the four distinct HWE-related tests discussed in Sections \ref{s:overview}, \ref{s:methods} and \ref{subsec:link}: the joint test of HWE and sdMAF, $T_{\mbox{\footnotesize RA, X-NPR$_{F\&M}$, }\hat{p}}$; the RA-based HWE test under the assumption of no sdMAF, $T_{\mbox{\footnotesize RA, X-NPR, }\hat{p}}$; the Pearson $\chi^2$-based HWE test under no sdMAF, $T_{\mbox{\footnotesize Pearson, X-NPR, }\hat{p}}$; and the female-only HWE test allowing sdMAF, $T_{\mbox{\footnotesize RA, X-NPR, }\hat{p}_f}$ as summarized in Table \ref{tab:pearson_tests}. For $T_{\mbox{\footnotesize Pearson, X-NPR, }\hat{p}}$, whose null distribution is a weighted sum of two independent $\chi^2_1$ variables, critical values were obtained empirically using $10^7$ random draws from $\chi^2_1 + \frac{m}{2f+m}\chi^2_1$.

For each simulation setting, we fixed the total sample size at $n=f+m=1000$, with varying female sample proportions of 25\%, 50\% or 75\% to examine the impact of sex imbalance. The male hemizygous genotypes were generated from $\text{Binomial}(m,p_m)$, while the female genotypes were generated from a multinomial distribution with probabilities determined by $p_f$ and $\delta_f$, respectively, the allele frequency and HWD magnitude in the females. Specifically, $p_{AA}=p_f^2+\delta_f$, $p_{Aa}=2p_f(1-p_f)-2\delta_f$, and $p_{aa}=(1-p_f)^2+\delta_f$, where $\delta_f=0$ corresponds to HWE in females. sdMAF is defined as $p_f-p_m$.

For the empirical assessment of T1E, we assumed HWE in all settings ($\delta_f=0$) and varied the frequency of the female minor allele from $p_f \in \{0.15, 0.20, \ldots, 0.50\}$. We considered sdMAF values in $\{0,\;0.05,\;0.10\}$, which correspond to no, moderate, and large sdMAF, respectively. The frequencies of male alleles were defined by $p_m=p_f-\text{sdMAF}$, ensuring that both $p_f$ and $p_m$ remained at least 0.05 in all scenarios. Thus, our simulations focus on common variants; rare variants are beyond the scope of this work.

Across simulation settings, we varied the magnitudes of sdMAF and HWD to assess robustness under model misspecification and power under alternative hypotheses. Each setting was replicated 10,000 times at the significance level $\alpha=0.05$. This test size is sufficient to reveal substantial T1E inflation when the no-sdMAF assumption is violated. For completeness, and to reflect more stringent GWAS quality-control settings, we further evaluated all four tests at $\alpha=10^{-4}$ using $10^6$ replicates; these results were qualitatively similar and are provided in {Supplementary Table S4}.

We primarily present results for non-negative sdMAF values. Due to symmetry, scenarios with negative sdMAF ($p_f<p_m$) yielded analogous results of equal magnitude in the opposite direction ({Supplementary Figure S1}). Additional simulations exploring a broader range of sdMAF values are presented in {Supplementary Figure S2}.

%%%%%%%%%%%%%%%%%%%%%%%%%%%%%%%%%%%%%%%%%%%%%%%%%%%%%%%%%%%%%%%%%%%%%%%%%%%%%%
%%%%%%%%%%%%%%%%%%%%%%%%%%%%%%%%%%%%%%%%%%%%%%%%%%%%%%%%%%%%%%%%%%%%%%%%%%%%%%
\subsection{Type I error rate for Xchr NPR SNPs} \label{subsubsec:t1e}

\begin{figure}[h]
    \centering
    \includegraphics{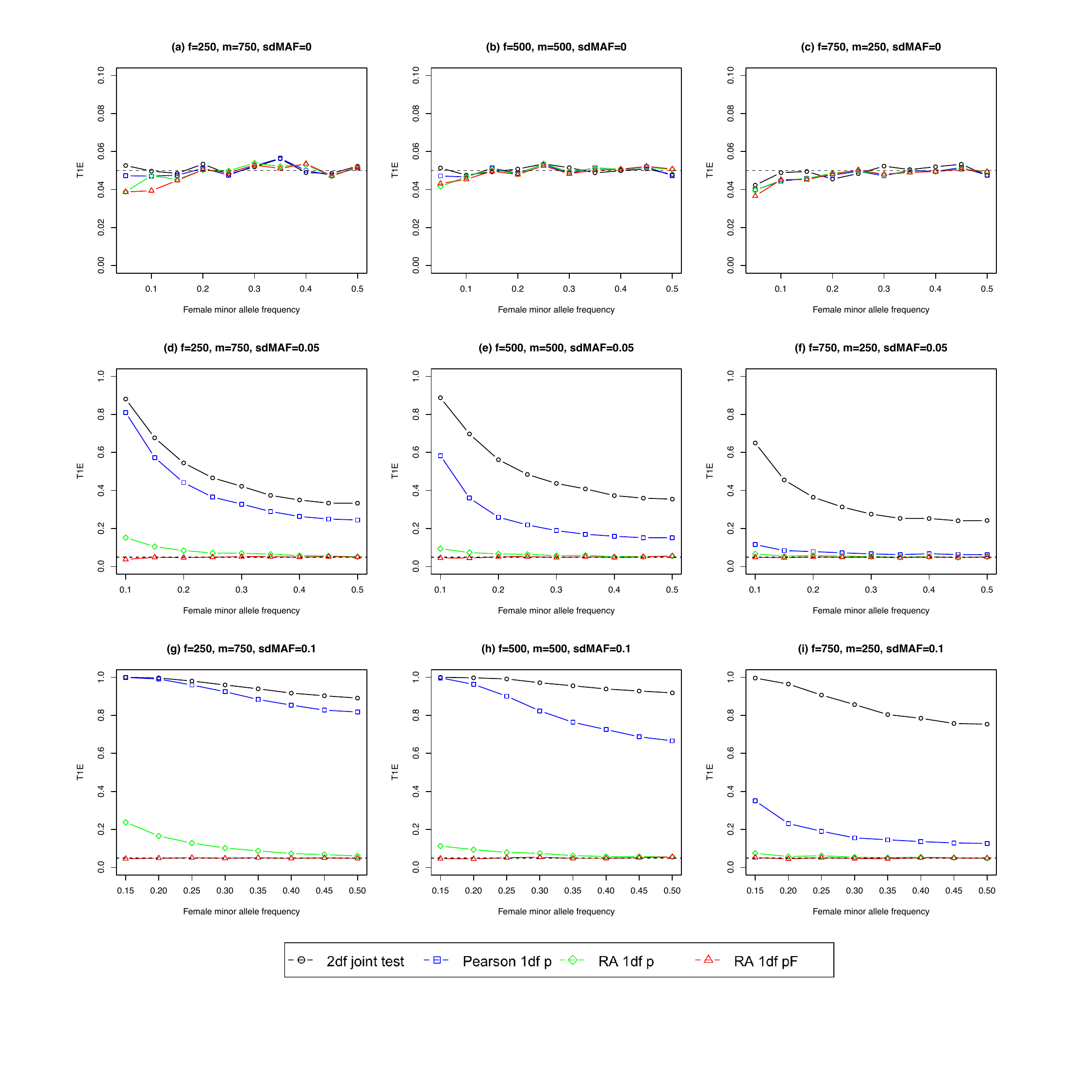}
    \vspace{-2.2cm}
    \caption{{\bf Empirical type I error rate of HWE tests for Xchr NPR SNPs}. The total sample size $n = 1,000$ with the female sample size $f = $ $250$, $500$ and $750$. The female minor allele frequency $p_f \in \{0.05, 0.50\}$ when sdMAF = $0$; $p_f \in \{0.10, 0.50\}$ when sdMAF = $0.05$; $p_f \in \{0.15, 0.50\}$ when sdMAF = $0.10$, all with $0.05$ increments. The $p_f$ starting value is determined to ensure $p_m > 0.05$. The female genotype frequencies follow the HWE assumption. The male allele frequency $p_m = p_f - \text{sdMAF}$, where sdMAF = $0$, $0.05$ and $0.1$. Each simulation scenario was repeated $10,000$ times, and the tests were evaluated using a significance level $\alpha = 0.05$. The black circles represent the 2 df joint tests of HWE and sdMAF; the blue squares represent the 1 df Pearson $\chi^2$ test under the assumption of no sdMAF; the green diamonds represent the 1 df RA regression-based HWE test under the assumption of no sdMAF; and the red triangles represent the 1 df HWE test under the assumption of sdMAF.}
    \label{fig:T1E}
\end{figure}

We first evaluated T1E under the null hypothesis of HWE in females ($\delta_f=0$) with varying levels of sdMAF and sex-specific sample compositions (Figure~\ref{fig:T1E}; {Supplementary Figures S9} for PAR SNPs). When sdMAF was absent ($p_f=p_m$), all four tests maintained appropriate T1E control at the nominal significance level, consistent with their theoretical null specifications.

However, as sdMAF increased, substantial differences emerged between the tests. The female-only test, $T_{\mbox{\footnotesize RA, X-NPR, }\hat{p}_f}$, remained robust at all levels of sdMAF, because it does not rely on sex-pooled allele frequency estimate or male genotype information. 

The RA-based test assuming that there is no sdMAF, $T_{\mbox{\footnotesize RA, X-NPR, }\hat{p}}$, exhibited moderate inflation of T1E as the sdMAF increased, reflecting the sensitivity to violations of its assumption of no sdMAF. In particular, because this test primarily captures the HWD component in the female samples, its inflation of T1E remained relatively modest when sdMAF was moderate (e.g., sdMAF $\leq 0.05$) or when males constituted a relatively small proportion of the study sample.

In contrast, the Pearson test with pooled allele frequency, $T_{\mbox{\footnotesize Pearson, X-NPR, }\hat{p}}$, showed substantially higher inflation of T1E, often severe even under moderate sdMAF, due to its implicit incorporation of sdMAF into the test statistic. In some settings, T1E approached 100\% at $\alpha=0.05$, demonstrating extreme vulnerability to misspecification of the assumption of no sdMAF. Only when sdMAF was moderate and female samples predominated did this test retain partial robustness. The magnitude of T1E distortion for tests assuming no sdMAF depended strongly on the sex composition of the sample. 

The joint test, $T_{\mbox{\footnotesize RA, X-NPR$_{F\&M}$, }\hat{p}}$, appropriately controls T1E only under its composite null hypothesis of HWE and no sdMAF.

\noindent{\bf Remark}. When sdMAF is present but male data are used, rejection by $T_{\text{\footnotesize RA, X-NPR$_{F\&M}$, }\hat{p}}$  reflects its sensitivity to sdMAF rather than evidence of HWD itself. Thus, interpretation of this joint test depends critically on whether the inferential goal is to detect HWD specifically or broader departures involving either HWD or sdMAF.

These findings empirically validate the theoretical relationships established in Section \ref{subsec:link}. In particular, the pronounced inflation of $T_{\mbox{\footnotesize Pearson, X-NPR, }\hat{p}}$ in T1E under sdMAF arises from its embedded sdMAF component, whereas RA-based tests provide more robust inference by explicitly separating HWD from sdMAF. Consequently, valid HWE inference on the X chromosome requires careful alignment between the chosen test and its underlying assumptions regarding sdMAF.

%%%%%%%%%%%%%%%%%%%%%%%%%%%%%%%%%%%%%%%%%%%%%%%%%%%%%%%%%%%%%%%%%%%%%%%%%%%%%%
%%%%%%%%%%%%%%%%%%%%%%%%%%%%%%%%%%%%%%%%%%%%%%%%%%%%%%%%%%%%%%%%%%%%%%%%%%%%%%
\subsection{Power for Xchr NPR SNPs}

Because the tests considered here differ in their corresponding null hypothesis, robustness to sdMAF, and inferential target, conventional power comparisons are only meaningful under carefully defined scenarios. Inflated rejection rates arising from model misspecification or broader composite null hypotheses may appear as increased power but do not necessarily reflect improved sensitivity to HWD itself. Consequently, power comparisons must be interpreted jointly with T1E behaviour and the specific scientific objective of the test. For power studies, we examined scenarios in which female genotypes departed from HWE ($\delta_f \neq 0$), with $\delta_f$ varying from $-0.04$ to $0.04$ with an increment of 0.005, and with multiple magnitudes of sdMAF and sample compositions of the two sexes (Figure~\ref{fig:power_no_sdMAF}; {Supplementary Figures S3--S8}).

\subsubsection{Power under no sdMAF}

\begin{figure}
    \centering
    \includegraphics{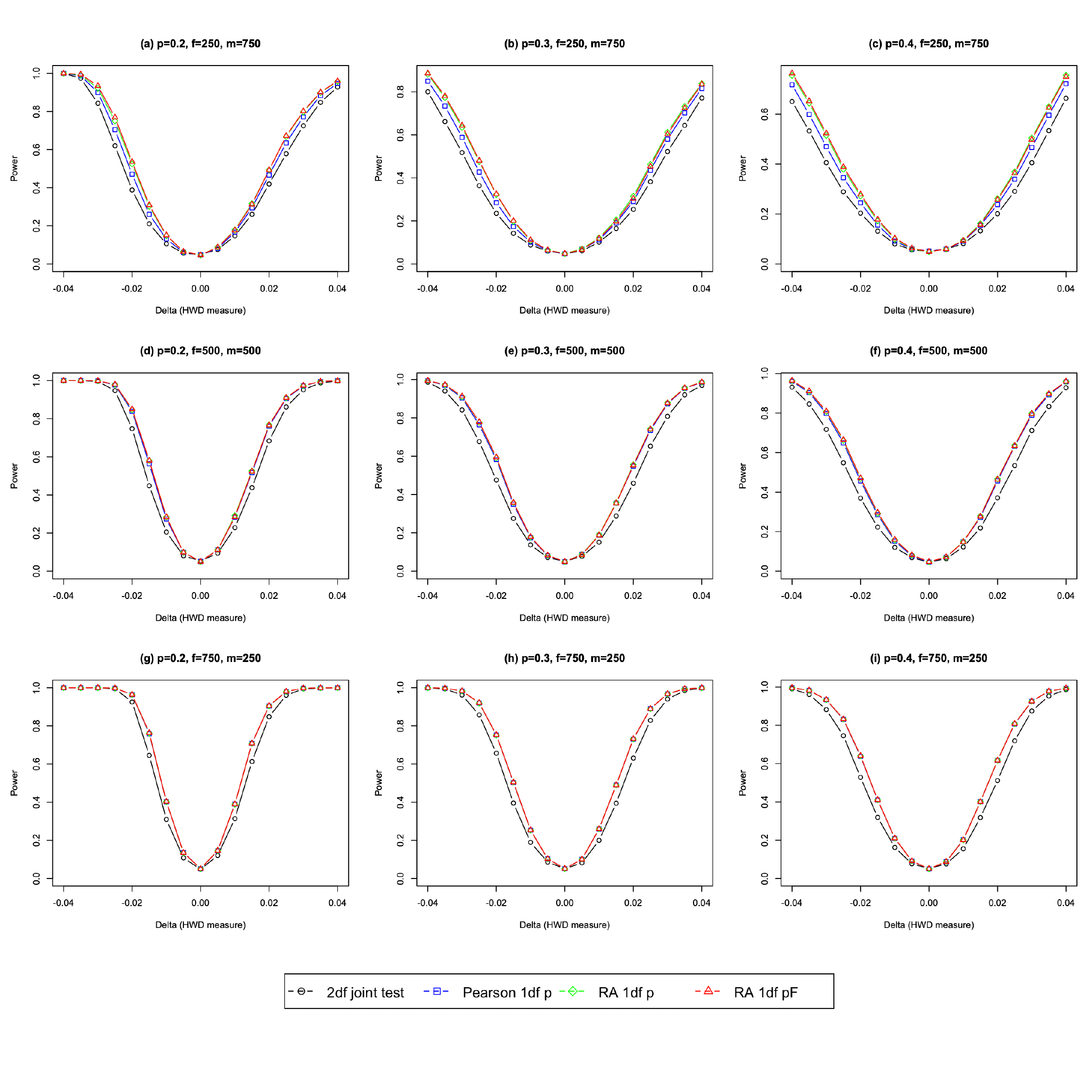}
    \vspace{-1.7cm}
    \caption{{\bf Empirical power of four HWE tests for Xchr NPR SNPs assuming no sdMAF}. The total sample size $n = 1,000$ with the female sample size $f = $ $250$, $500$ and $750$. The female and male minor allele frequencies are assumed identical, with $p_f = p_m = 0.2$, $0.3$ and $0.4$. The female genotype frequencies are $(p_{AA}, p_{Aa}, p_{aa})$, where $p_{AA}= p_f^2 + \delta_f$, $p_{Aa} = 2p_f(1-p_f) - 2 \delta$, and $p_{aa} =  (1-p_f)^2 - \delta_f $. The classical measure for HWD, $\delta_f$, takes values from \{$-0.04$, $0.04$\} with $0.005$ increment. Each simulation scenario was repeated $10,000$ times, and the tests were evaluated using a significance level $\alpha = 0.05$. The black circles represent the 2 df joint tests of HWE and sdMAF; the blue squares represent the 1 df Pearson $\chi^2$ test under the assumption of no sdMAF; the green diamonds represent the 1 df RA regression-based HWE test under the assumption of no sdMAF; and the red triangles represent the 1 df HWE test under the assumption of sdMAF.}
    \label{fig:power_no_sdMAF}
\end{figure}

We first considered the idealized setting of no sdMAF ($p_f=p_m$), in which all four tests exhibited appropriate T1E control (Section~\ref{subsubsec:t1e}). Figure~\ref{fig:power_no_sdMAF} summarizes the empirical powers across representative allele frequencies $p_f=p_m=\{0.2,0.3,0.4\}$.

Under the correct assumption of no sdMAF, the two RA-based HWE tests $T_{\mbox{\footnotesize RA, X-NPR, }\hat{p}_f}$ and $T_{\mbox{\footnotesize RA, X-NPR, }\hat{p}}$, consistently achieved the highest power in nearly all settings. In general, the power differences between these two tests were modest. A closer examination ({Supplementary Tables S1--S3}) shows that when $\delta_f<0$, $T_{\mbox{\footnotesize RA, X-NPR, }\hat{p}_f}$ tended to have slightly higher power than $T_{\mbox{\footnotesize RA, X-NPR, }\hat{p}}$, while the reverse was true when $\delta_f>0$. These finite-sample differences arise from the weighted $\widehat{\text{sdMAF}}^2$ term embedded in $T_{\mbox{\footnotesize RA, X-NPR, }\hat{p}}$, but the differences diminish asymptotically as the sample size increases or when the proportions of the male sample are relatively small.

The joint test, $T_{\mbox{\footnotesize RA, X-NPR}_{F\&M}, \hat{p}}$, was consistently less powerful than the two RA-based HWE tests because its composite null allocates degrees of freedom to both HWD and sdMAF components. The Pearson test with pooled allele frequency, $T_{\mbox{\footnotesize Pearson, X-NPR, }\hat{p}}$, also exhibited measurable but generally smaller power loss compared to the RA-based HWE tests, reflecting its partial weighting of the sdMAF component. Importantly, when the proportions of male samples were relatively low, this power loss became modest, and the Pearson test often approached the performance of the RA-based HWE procedures.

Overall, under the correctly specified no-sdMAF assumption, empirical power ordering closely matched the theoretical decompositions in Section \ref{subsec:link}: procedures that more directly target HWD while minimizing unnecessary sdMAF-related components achieve greater efficiency, while broader or partially composite tests incur predictable but generally modest finite-sample power costs.

%%%%%%%%%%%%%%%%%%%%%%%%%%%%%%%%%%%%%%%%%%%%%%%%%%%%%%%%%%%%%%%%%%%%%%%%%%%%%%
\subsubsection{Power considerations under sdMAF}

When sdMAF is present, formal power comparisons among the four tests become substantially less informative and may even be misleading. As shown in Section~\ref{subsubsec:t1e}, the two HWE tests that assume no sdMAF, namely $T_{\mbox{\footnotesize RA, X-NPR, }\hat{p}}$ and $T_{\mbox{\footnotesize Pearson, X-NPR, }\hat{p}}$, can exhibit a grossly inflated T1E under sdMAF and therefore are not appropriate for meaningful power evaluation in such settings.

Between the remaining two tests, $T_{\mbox{\footnotesize RA, X-NPR, }\hat{p}_f}$ and $T_{\mbox{\footnotesize RA, X-NPR$_{F\&M}$, }\hat{p}}$, direct power comparison is also inherently limited because the two tests address distinct null hypotheses. Specifically, $T_{\mbox{\footnotesize RA, X-NPR, }\hat{p}_f}$ evaluates HWD while allowing sdMAF. while $T_{\mbox{\footnotesize RA, X-NPR$_{F\&M}$, }\hat{p}}$ tests a broader composite null of both HWE and no sdMAF. 
Consequently, higher rejection rates for the joint test under sdMAF primarily reflect its broader composite alternative rather than increased sensitivity to HWD alone.

Thus, under sdMAF, the primary inferential message is not which test is ``more powerful'', but rather that the test choice must be aligned with the scientific objective. If the goal is to detect HWD specifically, $T_{\mbox{\footnotesize RA, X-NPR, }\hat{p}_f}$ provides the most appropriate and interpretable procedure. If simultaneous detection of either HWD or sdMAF is desired, then the joint test $T_{\mbox{\footnotesize RA, X-NPR}_{F\&M}, \hat{p}}$ may be preferable.

For completeness, we provide empirical rejection patterns under varying sdMAF and HWD scenarios in {Supplementary Figures S3--S8}. These results largely align with theoretical predictions: the joint test can exhibit higher rejection rates when HWD is weak because it captures sdMAF, whereas the female-only test remains focused solely on HWD. Consequently, these additional analyses primarily reinforce the broader conceptual framework rather than provide standalone power rankings.

\subsection{Comments on type I error and power for Xchr PAR SNPs}

We conducted supplementary simulation studies for Xchr PAR SNPs, with results presented in Supplementary Table S5 and Supplementary Figures S9–S11. However, in contrast to the NPR setting, the inferential behavior of PAR tests largely follows directly from well-established Pearson theory $\chi^2$ under the corresponding model assumptions because both females and males are diploids in PAR regions.

Specifically, when sdMAF is allowed, $T_{\mbox{\footnotesize RA, X-PAR, }\hat{p}_f,\ \hat{p}_m}$ is statistically equivalent to the sum of two independent sex-specific autosomal Pearson $\chi^2_1$ tests, analogous to standard HWE tests across multiple populations \citep{troendle1994note}. When no sdMAF is assumed, $T_{\mbox{\footnotesize RA, X-PAR, }\hat{p}}$ reduces to the classical pooled autosomal HWE test. Consequently, the type I error properties of the PAR tests follow directly from the standard Pearson $\chi^2$ theory with the correct model specification. 

For PAR SNPs, the principal inferential issue is generally not null calibration itself, but rather whether pooling across sexes is biologically and statistically appropriate. When allele frequencies and HWD patterns are similar between females and males, the pooled 1 df test is expected to be more efficient. In contrast, when sdMAF or sex-specific HWD is present, the 2 df test provides a more robust framework by explicitly preserving the sex-specific genotype structure.

The corresponding power relationships follow directly from the principles of classical hypothesis testing. Under homogeneous HWD and without sdMAF, the pooled 1 df procedure is generally more powerful because it targets a simpler alternative with fewer degrees of freedom. However, when HWD differs between sexes or sdMAF is substantial, the 2 df test may achieve superior sensitivity by avoiding model misspecification and information loss induced by inappropriate pooling.

Thus, rather than presenting largely redundant simulation results for statistically standard scenarios, we focus on the real-data PAR analyses in Section~\ref{s:app}, where these modeling distinctions have direct practical consequences for Xchr quality control. This approach preserves emphasis on the novel inferential challenges unique to Xchr while maintaining a unified framework across both the NPR and PAR settings.

%%%%%%%%%%%%%%%%%%%%%%%%%%%%%%%%%%%%%%%%%%%%%%%%%%%%%%%%%%%%%%%%%%%%%%%%%%%%%%
%%%%%%%%%%%%%%%%%%%%%%%%%%%%%%%%%%%%%%%%%%%%%%%%%%%%%%%%%%%%%%%%%%%%%%%%%%%%%%
\subsection{Application to the high-coverage 1000 Genomes Project (1kGP)}
\label{s:app}

To evaluate the practical implications of the proposed framework, we applied these Xchr HWE tests to the recently released high-coverage whole-genome sequencing (WGS) data from the expanded 1000 Genomes Project (1kGP) \citep{byrska2022high}. The project included participants from 26 populations across five super-populations: Africa (AFR), Americas (AMR), East Asia (EAS), Europe (EUR), and South Asia (SAS). {The high-coverage WGS data was aligned to the T2T-CHM13v2.0 human genome reference using XYalign, which has been shown to the quality of sex chromosome genotype calls \citep{rhie2023complete, webster2019identifying, garg2026assessing}.} Here, we focus on the AFR super-population to clarify the practical behaviour and inferential consequences of the proposed Xchr HWE tests. Results for the other super-populations are provided in {Supplementary Figures S12--S38} and show broadly consistent patterns.

We analyzed both Xchr NPR and PAR SNPs using unrelated individuals. The AFR super-population included 652 independent samples (336 females and 316 males). After standard quality-control (QC) procedures, the removal of up to second-degree relatives and restriction to common bi-allelic SNPs with non-missing genotypes (MAF $\geq 5\%$ in female, male and pooled samples), 240,165 NPR variants, 11,384 PAR1 variants, and 638 PAR2 variants remained for analysis. Detailed QC procedures are described in \citet{garg2026assessing}. Statistical significance was assessed using the conventional genome-wide significance threshold of $\alpha = 5 \times 10^{-8}$ \citep{marees2018tutorial, dudbridge2008estimation}.

%%%%%%%%%%%%%%%%%%%%%%%%%%%%%%%%%%%%%%%%%%%%%%%%%%%%%%%%%%%%%%%%%%%%%%%%%%%%%%
%%%%%%%%%%%%%%%%%%%%%%%%%%%%%%%%%%%%%%%%%%%%%%%%%%%%%%%%%%%%%%%%%%%%%%%%%%%%%%
\subsubsection{Xchr NPR regions}

\begin{figure}
    \centering
    \includegraphics{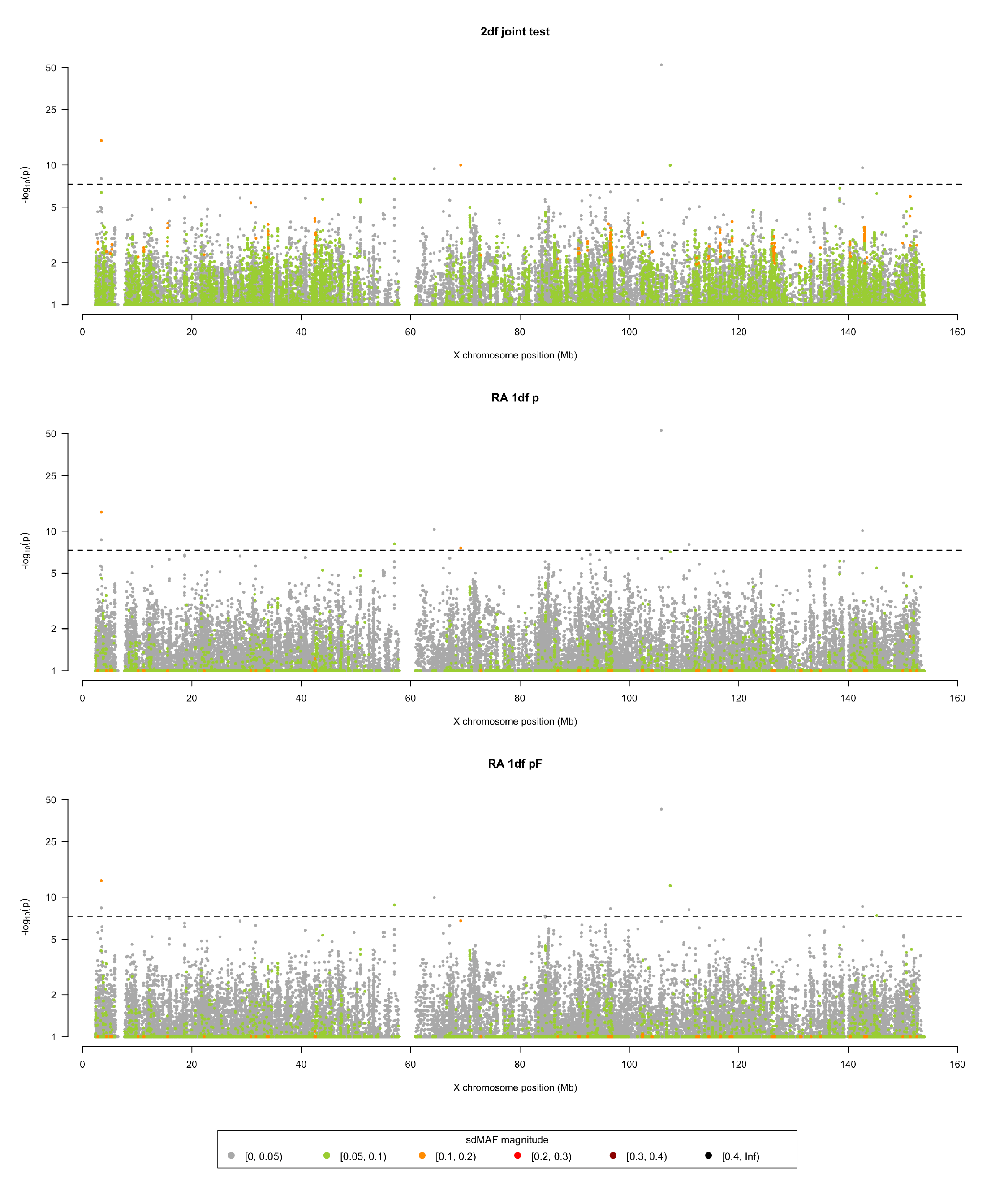}
    \caption{{\bf Manhattan plot of the HWE testing results in Xchr Non-Pseudoautosomal Region (NPR) of the AFR super-population}. Three RA-based HWE tests for NPR SNPs are compared here: the 2 df joint test of sdMAF and HWE ($T_{\text{\footnotesize RA, X-NPR$_{F\&M}$, }\hat{p}}$), the 1 df HWE test assuming no sdMAF ($T_{\text{\footnotesize RA, X-NPR, }\hat{p}}$), and the 1 df HWE test assuming sdMAF ($T_{\text{\footnotesize RA, X-NPR, }\hat{p}_f}$). The y-axis is $-\log_{10}$(p-values) and p-values $>0.1$ are plotted as $0.1$ ($1$ on $-\log_{10}$ scale) for better visualisation. The x-axis represents the genomic position of the variants in million base pairs (Mb) on the X chromosome. The colour of the dots represents the magnitude of sdMAF (the absolute value of sdMAF) in the corresponding variant. The dashed black line represents the genome-wide significance level $\alpha = 5 \times 10^{-8}$ ($7.3$ on $-\log_{10}$ scale). }
    \label{fig:Manhattan_NPR}
\end{figure}

Figure~\ref{fig:Manhattan_NPR} summarizes the test results across the Xchr NPR region. A central pattern is that variants with large sdMAF drive disproportionately significant results in the joint test, reflecting its broader composite null hypothesis. By contrast, the two 1 df HWE-focused procedures were substantially less influenced by the magnitude of sdMAF, thereby providing more targeted inference for HWD itself.

Pairwise comparisons ({Supplementary Figure S16}) further demonstrate that the joint test primarily gains additional rejection power when sdMAF is substantial, whereas the two 1 df HWE tests generally produce highly concordant results. This empirical behaviour closely mirrors the theoretical decompositions established in Section \ref{subsec:link}.

At the genome-wide significance threshold, Table~\ref{tab:sig_AFR_SNPs}(a) summarizes the 12 NPR SNPs that significantly deviated from HWE in at least one of the three RA-based tests. Seven SNPs were consistently identified by all three procedures, providing strong evidence of HWD regardless of the assumption of sdMAF. However, the remaining SNPs highlight the practical inferential consequences of differing null hypotheses and modeling assumptions.

In particular, two SNPs were identified primarily by the joint test and/or the HWE test assuming that there is no sdMAF in the presence of relatively large sdMAF. For example, rs765270695 (POS:69163175) exhibited substantial sdMAF ($|\widehat{\text{sdMAF}}|=0.116$), suggesting that its significance in the joint test was driven in part by sdMAF itself. Although $T_{\text{\footnotesize RA, X-NPR, }\hat{p}}$ also detected this SNP, this significance may partly reflect the inflation of T1E under a large sdMAF, consistent with our simulation findings. Similarly, rs1017422089 (POS:107457632), with an sdMAF magnitude of 0.08, further clarifies how substantial sdMAF can materially influence rejection patterns across tests.

In contrast, three SNPs were uniquely identified by the female-only test $T_{\text{\footnotesize RA, X-NPR, }\hat{p}_f}$. These variants generally exhibited modest absolute HWD and only mild sdMAF, but relatively large, scaled female-specific disequilibrium ($\hat{\rho}_f$), demonstrating the improved sensitivity of the female-only procedure when female MAF is correctly specified.

Overall, these empirical findings demonstrate that test choice can materially alter variant prioritization in real genomic analyses. Procedures that conflate sdMAF with HWD may preferentially identify variants driven partly by sex-specific allele-frequency differences, whereas the female-only framework more directly targets female-specific HWD. Thus, the RA framework improves interpretability by explicitly separating HWD from sdMAF, thereby reducing the risk of conflating these distinct sources of deviation in Xchr quality control through HWE testing.

\begin{landscape}
% \newgeometry{left=6cm, top=13cm}

\begin{table}[h]
\centering \scriptsize
\begin{tabular}{l | r | c | r | r | c| c | c | r || c| c | c  | c }
\hline \hline      
\multicolumn{1}{c|}{rs number}  & \multicolumn{1}{c|}{Position} & \multicolumn{1}{c|}{Ref/Alt} & \multicolumn{1}{c|}{$\hat{\delta}_f$} & \multicolumn{1}{c|}{$\hat{\delta}_m$} &  $\hat{p}_f$  &  $\hat{p}_m$ & $\hat{p}$  &  \multicolumn{1}{c||}{$\hat{p}_f - \hat{p}_m $} &  \multicolumn{4}{c}{p-values}   \\ \hline \hline
\multicolumn{9}{c ||}{\bf (a) NPR SNPs} & \multicolumn{1}{c|}{$T_{{\text{sdMAF}}}$}  & $T_{\text{\scriptsize RA, X-NPR$_{F\&M}$, }\hat{p}}$  & $T_{\text{\scriptsize RA, X-NPR, }\hat{p}}$ & $T_{\text{\scriptsize RA, X-NPR, }\hat{p}_f}$ \\ \hline
rs6655837 & 3448664 & A/G & -0.098 & \multicolumn{1}{c|}{--} & 0.405  & 0.297 & 0.370 & 0.107 & 2.83e-04 & \textbf{1.07e-15} & \textbf{2.21e-14} & \textbf{7.26e-14} \\
rs1278131 &   3455479 & A/C & -0.067 & \multicolumn{1}{c|}{--} & 0.296 & 0.266 & 0.286  & 0.030 & 0.292 & \textbf{1.01e-08} & \textbf{2.13e-09} & \textbf{4.11e-09} \\
rs6612851 &  57025923 & C/T & -0.065 & \multicolumn{1}{c|}{--} & 0.272  & 0.329 & 0.290  & -0.057 & 0.058 & \textbf{1.08e-08} & \textbf{7.83e-09} &  \textbf{1.60e-09} \\
rs7879488 &  64320997 & A/G &  0.024 & \multicolumn{1}{c|}{--} & 0.074 & 0.070 & 0.073  & 0.005 & 0.796 & \textbf{4.00e-10} & \textbf{4.92e-11} & \textbf{1.13e-10} \\
N/A & 105854847 & T/C &  0.087 & \multicolumn{1}{c|}{--} & 0.132  & 0.085 &  0.117 & 0.047  & 0.045 & \textbf{5.63e-53} & \textbf{2.88e-53} & \textbf{1.94e-43} \\
rs28788859 & 110914049 & G/A  & -0.078 & \multicolumn{1}{c|}{--} & 0.467 & 0.513 & 0.482  & -0.045 & 0.160 & \textbf{2.81e-08} & \textbf{9.22e-09} &  \textbf{7.60e-09} \\
rs859902 & 142631953 & T/C & 0.027 & \multicolumn{1}{c|}{--} & 0.092 & 0.066 & 0.084 & 0.026 & 0.175 & \textbf{2.71e-10} & \textbf{8.25e-11} & \textbf{2.62e-09} \\ \hline
rs5936969 & 69163175 & C/T & -0.059 & \multicolumn{1}{c|}{--} & 0.293 & 0.177 & 0.256 &  0.116 & 8.98e-06 & \textbf{9.88e-11} &  \textbf{2.70e-08} & 1.67e-07 \\
rs764585100 &  107457632 & A/T &  0.023 & \multicolumn{1}{c|}{--} & 0.062 & 0.142 & 0.088 & -0.080 & 3.91e-04 & \textbf{1.06e-10} &  7.80e-08 & \textbf{8.21e-13} \\ \hline
rs5968817 &  84588702 & A/C &  0.017 & \multicolumn{1}{c|}{--} & 0.061 & 0.085 & 0.069 & -0.024 & 0.197 & 2.12e-06 & 9.04e-07 & \textbf{4.38e-08} \\
rs73550265 &  96543648 & T/C &  0.015 & \multicolumn{1}{c|}{--} & 0.051 & 0.066 & 0.056 & -0.016 & 0.352 & 3.78e-07 & 9.14e-08 &  \textbf{5.30e-09} \\
rs150556780 & 145219770 & G/A & 0.021 & \multicolumn{1}{c|}{--} & 0.076 & 0.130 & 0.093 & -0.054 & 0.015 & 5.56e-07 & 3.69e-06 &  \textbf{3.90e-08} \\
\hline \hline
  \multicolumn{9}{c||}{\bf (b) PAR1 SNPs} &\multicolumn{1}{c|}{$T_{{\text{sdMAF}}}$}  & $T_{\text{\scriptsize RA, X-PAR, }\hat{p}}$ & $T_{\text{\scriptsize RA, X-PAR, }\hat{p}_f, \: \hat{p}_m}$ & \multicolumn{1}{c}{--} \\ \hline
rs867436760 &  11391 & A/G & -0.119 & -0.108  & 0.382 & 0.386 & 0.384 & -0.004  & 0.852 & \textbf{1.03e-34} & \textbf{1.27e-33} & \multicolumn{1}{c}{--} \\
rs73178918 & 1184574 & C/G & -0.097 & -0.103  & 0.360 & 0.354  & 0.357 & 0.006  & 0.775 & \textbf{8.99e-29} & \textbf{1.13e-27} & \multicolumn{1}{c}{--} \\ 
rs184807393 & 249017  & A/G & 0.049  & 0.035  & 0.247 & 0.282 & 0.264 & -0.035 & 0.198 & \textbf{2.54e-08} & 8.12e-08 & \multicolumn{1}{c}{--} \\ 
rs2259750 & 2387607 & G/A  & 0.002 & -0.089 & 0.113 & 0.460  & 0.281 & -0.347 & 1.05e-66 & 1.39e-01 & \textbf{1.75e-09} & \multicolumn{1}{c}{--} \\
rs2857317 & 2393813 & A/G & 0.001 & -0.179  & 0.119 & 0.547  & 0.327 & -0.428 & 1.47e-152 & 2.21e-06 & \textbf{9.74e-37} & \multicolumn{1}{c}{--} \\
\hline \hline
\multicolumn{9}{c ||}{\bf (c) PAR2 SNPs} &\multicolumn{1}{c|}{$T_{{\text{sdMAF}}}$}  & $T_{\text{\scriptsize RA, X-PAR, }\hat{p}}$ & $T_{\text{\scriptsize RA, X-PAR, }\hat{p}_f, \: \hat{p}_m}$ & \multicolumn{1}{c}{--} \\ \hline
rs306932 & 153946131 & C/T &  0.004 & -0.151  & 0.177 & 0.576  & 0.370 & -0.399 & 1.97e-95 & 5.94e-04 & \textbf{3.87e-27} & \multicolumn{1}{c}{--} \\
rs306921 & 153949768 & A/T &  0.012 & -0.131 & 0.268 & 0.601  & 0.429 & -0.333 & 4.27e-52 & 2.10e-03 & \textbf{2.58e-21} & \multicolumn{1}{c}{--} \\
rs306903 & 153964583 & C/T & -0.007 & -0.133  & 0.240 & 0.598  & 0.413 & -0.359 & 7.52e-67 & 1.58e-04 & \textbf{7.55e-22} & \multicolumn{1}{c}{--} \\
rs306898 & 153972806 & C/T & -0.001 & -0.100 & 0.333 & 0.650  & 0.487 & -0.317 & 4.37e-43 & 1.44e-02 & \textbf{4.96e-14} & \multicolumn{1}{c}{--} \\
\hline \hline
\end{tabular}

\caption{{\bf The Xchr NPR and PAR SNPs that significantly deviate from HWE in the AFR super-population of 1kGP}. The HWD measure are reported in the female sample ($\hat{\delta}_f$) and in the male sample ($\hat{\delta}_m$), respectively; the sample allele frequency estimates ($\hat{p}_f$, $\hat{p}_m$) measures the frequency of the female minor allele in female and male sample, respectively; the sample sdMAF ($\widehat{\text{sdMAF}} = \hat{p}_f - \hat{p}_m$) and the sdMAF test p-values ($T_{{\text{sdMAF}}}$) are also reported to facilitate result interpretations. {\bf (a) NPR SNPs}. Each SNP is significant at $\alpha = 5 \times 10^{-8}$ by at least one of the joint test ($T_{\text{\footnotesize RA, X-NPR$_{F\&M}$, }\hat{p}}$), the HWE test assuming no sdMAF ($T_{\text{\footnotesize RA, X-NPR, }\hat{p}}$), and the HWE test assuming sdMAF ($T_{\text{\footnotesize RA, X-NPR, }\hat{p}_f}$). The rows are organized as follows: exemplary SNPs where all three tests are significant; SNPs where at least one test is {\it not} significant; and SNPs where only the HWE test assuming sdMAF $T_{\text{\footnotesize RA, X-NPR, }\hat{p}_f}$ is significant. The normalized allele intensity plots of these significant SNPs are presented in {Supplementary Figures S40}. {\bf (b) PAR1 SNPs and (c) PAR2 SNPs}. Each SNP is significant at $\alpha = 5 \times 10^{-8}$ by at least one of the 1 df HWE test $T_{\text{\footnotesize RA, X-PAR, }\hat{p}}$ assuming no sdMAF, and the 2 df HWE test $T_{\text{\footnotesize RA, X-PAR, }\hat{p}_f, \: \hat{p}_m}$ assuming sdMAF. The rows are organized as: exemplary SNPs where both tests are significant; only the test assuming no sdMAF ($T_{\text{\footnotesize RA, X-PAR, }\hat{p}}$) is significant; only the test assuming sdMAF ($T_{\text{\footnotesize RA, X-PAR, }\hat{p}_f, \: \hat{p}_m}$) is significant. See {Supplementary Tables S16-19} for the full list of PAR1 and PAR2 SNPs that significantly deviate from HWE. The normalized allele intensity plots of these significant SNPs are presented in {Supplementary Figures S41-42}.}
    \label{tab:sig_AFR_SNPs}
\end{table}

\end{landscape}

%%%%%%%%%%%%%%%%%%%%%%%%%%%%%%%%%%%%%%%%%%%%%%%%%%%%%%%%%%%%%%%%%%%%%%%%%%%%%%
%%%%%%%%%%%%%%%%%%%%%%%%%%%%%%%%%%%%%%%%%%%%%%%%%%%%%%%%%%%%%%%%%%%%%%%%%%%%%%
\subsubsection{Xchr PAR regions}

% Move to supp: Figure~\ref{fig:pp_PAR1}, Figure~\ref{fig:pp_PAR2}, 

\begin{figure}
    \centering
    \includegraphics{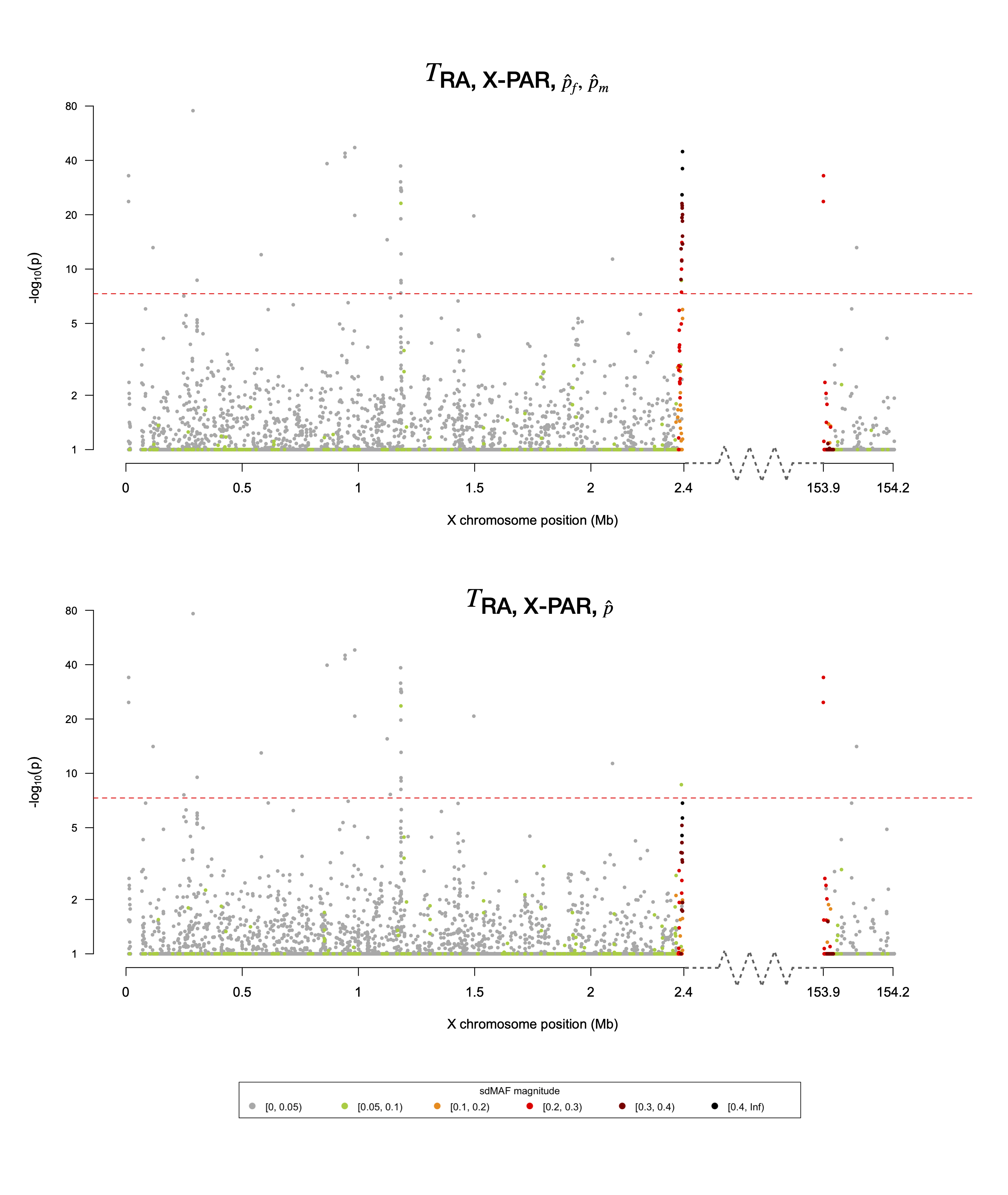}
     \vspace{-1.5cm}
    \caption{{\bf Manhattan plot of the HWE testing results in Xchr PAR1 and PAR2 of the AFR super-population}. Two RA-based HWE tests for PAR SNPs are compared: the 2 df test assuming sdMAF ($ T_{\text{\footnotesize RA, X-PAR, }\hat{p}_f, \: \hat{p}_m}$), the 1 df test assuming no sdMAF ($ T_{\text{\footnotesize RA, X-PAR, }\hat{p}}$). The y-axis is $-\log_{10}$(p-values) and p-values $>0.1$ are plotted as $0.1$ ($1$ on $-\log_{10}$ scale) for better visualisation. The x-axis represents the genomic position of the variants in million base pairs (Mb) on the X chromosome. The colour of the dots represents the magnitude of sdMAF (the absolute value of sdMAF) in the corresponding variant. The dashed black line represents the genome-wide significance level $\alpha = 5 \times 10^{-8}$ ($7.3$ on $-\log_{10}$ scale). }
    \label{fig:Manhattan_PAR}
\end{figure}

We next evaluated HWE inference in the two Xchr PAR regions: PAR1 (POS: 0--2.4Mb) at the tips of the short arms of the X and Y chromosomes and PAR2 (POS: 153.9--154.2Mb) at the tips of the long arms. Although both females and males are diploids in PAR regions, substantial sdMAF may still arise, particularly near the PAR--NPR boundaries \citep{wang_major_2022}. In the AFR super-population, sdMAF magnitudes in the PAR regions were often considerably larger than those observed in the NPR regions, with maximum sdMAF magnitudes reaching 0.47 in PAR1 and 0.40 in PAR2. In particular, SNPs with large sdMAF clustered near the PAR boundary regions (Figure \ref{fig:Manhattan_PAR}).

For PAR SNPs, we compared two HWE tests: a 1 df test assuming that there is no sdMAF, $T_{\text{\footnotesize RA, X-PAR, }\hat{p}}$ ({Supplementary Notes 3.1.2}), and a 2 df test allowing for sdMAF, $T_{\text{\footnotesize RA, X-PAR, }\hat{p}_f, \: \hat{p}_m}$ (\ref{eq:par_sdmaf}), which is equivalent to the sum of female- and male-specific Pearson $\chi^2_1$ tests. Importantly, unlike the joint NPR test, both PAR procedures focus exclusively on HWE inference; thus, the large sdMAF itself is not a direct source of power for HWE testing. 

In both PAR1 and PAR2, the two tests generally produced highly concordant results when sdMAF was mild. However, near PAR boundaries where the sdMAF was substantial, the 2 df test frequently identified strong HWD signals that were partially or entirely missed by the simpler pooled 1 df procedure. This pattern was especially pronounced near the PAR1--NPR and PAR2--NPR boundaries, where sex-specific genotype structures often differ markedly.

Detailed examination of significant PAR SNPs (Table~\ref{tab:sig_AFR_SNPs} (b)-(c); {Supplementary Tables S16--S19}) further reinforced these findings. Variants with large sdMAF near boundary regions frequently exhibited female genotypes close to HWE but substantial male-specific HWD, often characterized by pronounced excess heterozygosity in males (negative $\hat{\delta}_m$). Under the sex-pooled no-sdMAF assumption, these male-specific disequilibrium patterns were often obscured by combined estimation, leading to false negatives for $T_{\text{\footnotesize RA, X-PAR, }\hat{p}}$. In contrast, the 2 df framework preserved sensitivity by explicitly modeling female and male HWD separately.

Overall, these results demonstrate that even in PAR regions, where the genotype structure superficially resembles autosomes, explicit modeling of sdMAF can materially improve HWE inference. The RA framework naturally extends to these settings by preserving interpretability across distinct genomic architectures, providing a unified statistical foundation for Xchr-wide HWE quality control.

%%%%%%%%%%%%%%%%%%%%%%%%%%%%%%%%%%%%%%%%%%%%%%%%%%%%%%%%%%%%%%%%%%%%%%%%%%%%%%
%%%%%%%%%%%%%%%%%%%%%%%%%%%%%%%%%%%%%%%%%%%%%%%%%%%%%%%%%%%%%%%%%%%%%%%%%%%%%%
%%%%%%%%%%%%%%%%%%%%%%%%%%%%%%%%%%%%%%%%%%%%%%%%%%%%%%%%%%%%%%%%%%%%%%%%%%%%%%
\section{Discussion}
\label{s:discussion} 

We developed a general, unifying statistical framework for Hardy--Weinberg equilibrium inference on the X chromosome using the robust allele-based regression model. By explicitly parameterizing HWD and sdMAF within a common regression structure, this framework clarifies that existing Xchr HWE tests are not merely alternative implementations of a single inferential procedure, but rather correspond to distinct null hypotheses under differing biological and statistical assumptions. This perspective resolves a longstanding ambiguity in Xchr HWE testing, unifies classical Pearson-type procedures within a broader inferential framework, and provides a principled basis for selecting appropriate tests for specific scientific objectives. Thus, rather than viewing Xchr HWE testing as a single universal problem, our framework emphasizes that appropriate inference requires explicit matching between scientific objective, biological plausibility, and statistical model.

A particularly important contribution of this work is the reinterpretation of widely used Pearson $\chi^2$-based Xchr tests within the RA framework. For example, the pooled Pearson test for Xchr NPR SNPs proposed by \citet{graffelman2016testing}, originally motivated as an HWE test, can be theoretically re-expressed in terms of its RA equivalent and shown to implicitly function as a joint test of HWE and sdMAF. This previously under-recognized property explains its sensitivity to sdMAF and potential inflation of type I error when sdMAF is present. More broadly, the RA framework demonstrates that many existing Xchr HWE procedures can be understood as structured score tests under explicit parameter constraints, thereby substantially improving interpretability and inferential transparency.

For practitioners, our results offer several practical recommendations. For Xchr NPR SNPs, when sdMAF is plausible, unknown, or itself of potential quality-control concern, the female-only RA-based HWE test $T_{\mbox{\footnotesize RA, X-NPR, }\hat{p}_f}$ generally provides the most robust and interpretable choice, because it directly targets female HWD without relying on potentially invalid pooling assumptions. When strong prior evidence supports the absence of sdMAF, pooled 1 d.f. procedures such as $T_{\mbox{\footnotesize RA, X-NPR, }\hat{p}}$ may offer modest efficiency gains. Joint tests such as $T_{\mbox{\footnotesize RA, X-NPR$_{F\&M}$, }\hat{p}}$ are best reserved for settings where simultaneous detection of HWD or sdMAF is scientifically relevant, rather than for routine HWE screening. For Xchr PAR SNPs, sex-pooled 1 df tests are appropriate when allele frequencies and HWD are reasonably homogeneous between sexes, whereas sex-stratified 2 df procedures are preferable near PAR boundaries or whenever substantial sdMAF is biologically plausible or empirically observed.

Our simulation studies and empirical analyses demonstrate that these distinctions are not simply theoretical. In real genomic data, inappropriate assumptions regarding sdMAF can substantially alter type I error, power, and variant prioritization. Tests that implicitly conflate sdMAF with HWD can yield false positives, false negatives, or misleading conclusions, particularly in large-scale quality-control applications. At the same time, sdMAF itself may arise from technical artifacts, biological processes, population structure, or genomic architecture. Explicitly separating sdMAF from HWD therefore improves both statistical rigor and practical interpretability, reducing the risk of conflating distinct sources of deviation in Xchr quality control.

Beyond HWE testing, the RA Xchr framework provides a versatile foundation for broader allele-based modeling on the X chromosome. Because the regression structure naturally accommodates covariate adjustment, multiple heterogeneous populations, and sex-specific genotype architectures, it can be extended to sdMAF testing under various assumptions ({Supplementary Notes 4}), as well as future applications involving association testing for sexually dimorphic phenotypes, related individuals, admixed populations, and other complex genomic study designs. More broadly, this framework provides an important methodological basis for generalized Xchr statistical inference.

Several limitations warrant consideration. First, our primary focus was on common variants and asymptotic score-test behaviour; rare variants or small-sample settings may require exact, permutation-based, or alternative specialized procedures. Second, while the framework clarifies null specification, practical interpretation of significant sdMAF remains context-dependent and may require further biological or technical investigation. Third, although our simulations focused primarily on NPR SNPs where inferential complexity is greatest, additional evaluation in specialized genomic contexts may further refine practical recommendations. Finally, an important future direction is to extend the framework to related samples. Although the RA formulation naturally suggests covariance-based generalizations, X-chromosome relatedness introduces additional complexity because the covariance structure depends on both kinship and sex-specific inheritance patterns. Therefore, extension of HWE inference to related samples on the Xchr requires additional methodological development beyond the scope of the present work.

In summary, this work establishes a unified and interpretable statistical foundation for HWE inference across Xchr NPR and PAR regions. By clarifying the different assumptions and inferential targets underlying existing procedures, explicitly separating HWD from sdMAF, and naturally expanding to broader regression-based applications, the RA framework substantially improves the rigour, flexibility, and interpretability of Xchr genetic quality-control analyses. As genomic studies increasingly incorporate the X chromosome into large-scale association and sequencing analyses, such principled statistical foundations will be essential for valid, reproducible, and scientifically meaningful inference.
\newpage

\section{Acknowledgment}
This research is funded by the Natural Sciences and Engineering Research Council of Canada (NSERC, RGPIN-04934 and RGPAS-522594), and the Canadian Institutes of Health Research (CIHR, MOP-310732) to LS. We would like to thank Elika Garg for pre-processing the 1000 Genomes Project data. Generative AI tools (ChatGPT, OpenAI) were used to edit language and refine the stylistic quality of a fully drafted manuscript. All scientific content, statistical methodology, interpretations, and final editorial decisions were determined and verified by the authors. ChatGPT Pro was used as a copy editor on a fully drafted manuscript.

\section{Data Availability Statement}
The high-coverage 1000 Genome Project (1kGP) WGS data is publicly avaiable at \url{https://s3-us-west-2.amazonaws.com/human-pangenomics/index.html?prefix=T2T/CHM13/assemblies/variants/1000_Genomes_Project/chm13v2.0/unrelated_samples_2504/}. 

\bibliographystyle{ametsoc2014}
\bibliography{Xchr_HWE.bib}

\end{document}